\documentclass[preprint,prd,aps,tighten,nofootinbib,amssymb]{revtex4}
\pdfoutput=1
\usepackage{graphicx}
\usepackage{epsfig,amssymb,amsmath}
\usepackage{slashed, color}
\usepackage{hyperref}
\usepackage{gensymb}
\usepackage[normalem]{ulem}

\usepackage{color} \usepackage{cancel}
\usepackage{float} \usepackage{graphicx}
\usepackage{subfigure} \usepackage{caption}	
\usepackage{dcolumn}
\usepackage{tikz}

\newcommand{\bea}{\begin{eqnarray}} \newcommand{\ena}{\end{eqnarray}}
\newcommand{\beq}{\begin{equation}} \newcommand{\eeq}{\end{equation}}

           \renewcommand{\l}{\lambda}
                
           \renewcommand{\O}{\Omega}

            \renewcommand{\t}{\tau}

\newcommand{\tikzcircle}[2][green,fill=green]{\tikz[baseline=-0.5ex]\draw[#1,radius=#2] (0,0) circle ;}

\newcommand{\lsim}{
\mathrel{\hbox{\rlap{\hbox{\lower4pt\hbox{$\sim$}}}\hbox{$<$}}}}
\newcommand{\gsim}{
\mathrel{\hbox{\rlap{\hbox{\lower4pt\hbox{$\sim$}}}\hbox{$>$}}}}

\begin{document}
\draft
\tighten
\preprint{CTPU-PTC-18-20}

\title{\large \bf Constraints on the reheating parameters after Gauss-Bonnet inflation from primordial gravitational waves }

\author{
    Seoktae Koh$^{a}$\footnote{Electronic address: kundol.koh@jejunu.ac.kr},
    Bum-Hoon Lee$^{b}$\footnote{Electronic address: bhl@sogang.ac.kr},
    Gansukh Tumurtushaa$^{c}$\footnote{Electronic address: gansuhmgl@ibs.re.kr},}
     
\affiliation{
 $^a$Department of Science Education, Jeju National University, Jeju, 63243, Korea\\
 $^b$Center for Quantum Spacetime, Sogang University, Seoul 121-742, Korea \\ Department of Physics, Sogang University, Seoul 121-742, Korea\\
 $^c$Center for Theoretical Physics of the Universe,  Institute for Basic Science, Daejeon 34051, Korea
    }

\vspace{4cm}

\begin{abstract}

We study the effects of  the Gauss-Bonnet term on the energy spectrum of inflationary gravitational waves. The models of inflation are classified into two types based on their predictions for the tensor power spectrum: red-tilted ($n_T<0$) and blue-tilted spectra ($n_T>0$), respectively, and then the energy spectra of the gravitational waves are calculated for each type of model. We find that the gravitational wave spectra are enhanced depending on the model parameter if the predicted inflationary tensor spectra have a blue tilt, whereas they are suppressed for the spectra that have a red tilt. Moreover, we perform the analyses on the reheating parameters involving the temperature, the equation-of-state parameter, and the number of $e$-folds using the gravitational wave spectrum. Our results imply that the Gauss-Bonnet term plays an important role not only during inflation but also during reheating whether the process is instantaneous or lasts for a certain number of $e$-folds until it thermalizes and eventually completes.

\end{abstract}

\maketitle
\section{Introduction}
Cosmic inflation~\cite{Guth:1980zm, Albrecht:1982wi, Linde:1981mu} is widely believed to be a successful paradigm for the early Universe that solves major problems in standard big bang cosmology.  It also predicts the scale-invariant spectrum of the anisotropies of the cosmic microwave background (CMB) and provides the seeds to the large-scale structure of the Universe~\cite{Hinshaw:2012aka, Komatsu:2010fb, Komatsu:2008hk, Planck:2013jfk, Ade:2015lrj}. Moreover, inflation predicts the generation of the primordial gravitational wave (PGW), the ripples in the curvature of spacetime. The existence of the PGW background can be confirmed indirectly by the detection of the B-mode CMB polarization, which is induced by the tensor fluctuation modes~\cite{Starobinsky:1979ty, Allen:1987bk,Sahni:1990tx, Kamionkowski:1996ks}, and directly by the ongoing and future mission concepts of the ground- and space-based laser interferometric detectors and the pulsar timing experiments~\cite{AmaroSeoane:2012km, Ferdman:2010xq, Kudoh:2005as}.

The temperature of the Universe during the period of inflation became almost zero; hence, it is necessary to reheat the Universe after inflation came to the end. In order to reheat the Universe, the inflaton field is considered to be oscillating around the minimum of its potential, and it transfers its energy to a plasma of the standard model particles. This period, a transition era between the end of inflation and the beginning of the radiation-dominated era, is known as the reheating epoch. Because no direct cosmological observations are traceable from this period of reheating, the physics of reheating is highly uncertain and unconstrained. Thus, this era depends heavily on models of inflation.

The Universe is transparent to the {\color{black} gravitational waves} up to the Planck era. The detection of the PGW background by a future observation would open up a new window in exploring the early Universe, particularly, the reheating era. It was also claimed that the temperature of reheating and the equation-of-state parameter during reheating can be probed by looking at the spectrum of the GW background~\cite{Turner:1993vb, Zhao:2006mm, Watanabe:2006qe, Nakayama:2008wy, Kuroyanagi:2008ye, Giovannini:2008tm, Nakayama:2009ce, Kuroyanagi:2014nba, Tumurtushaa:2016ars, Koh:2015brl, Kuroyanagi:2011fy, Buchmuller:2013lra}. Therefore, in this work, we consider inflationary models with a Gauss-Bonnet (GB) term to estimate the energy spectrum of the  PGW and to provide constraints on the reheating parameters. Inflationary models with a GB term are not uncommon, and it is well studied in the context of inflation, dark energy, and the PGW~\cite{Nojiri:2005vv, Guo:2010jr, Satoh:2008ck, Satoh:2010ep, Chakraborty:2018scm, Koh:2014bka, Jiang:2013gza, Koh:2016abf, Satoh:2007gn}, as well as for reheating~\cite{vandeBruck:2016xvt, Bhattacharjee:2016ohe, Nozari:2017rta}.

Following the approach proposed in Refs.~\cite{Dai:2014jja, Creminelli:2014fca, Munoz:2014eqa, Cai:2015soa}, we perform the analyses on the reheating parameters including the equation of state, the duration, and the temperature of reheating. Since the  reheating parameters are often linked to the observable quantities of inflation such as the scalar and tensor spectral indices, their running spectral indices, the tensor-to-scalar ratio, and the number of $e$-folds during inflation, one can provide constraints on the model parameters in light of current and future observation~\cite{Hinshaw:2012aka, Komatsu:2010fb, Komatsu:2008hk, Planck:2013jfk, Ade:2015lrj, Amendola:2012ys, Andre:2013afa}.

The paper is organized as follows. In Sec.~\ref{sec:GBI}, we review the basics of inflationary models with a GB term and the observable quantities. We classify inflationary models with a GB term into two types in Sec.~\ref{sec:class}; models that predict the inflationary tensor power spectrum with a red tilt and those with a blue tilt, respectively. With these models, we calculate the energy spectrum of the PGW in Sec.~\ref{sec:GWES}.
Motivated by the fact that the reheating temperature can be determined by the detection of the PGW background, we further perform the analyses on the reheating parameters and provide constraints on those parameters in Sec.~\ref{sec:reh}. Finally, the summary and the conclusion of the present work are provided in Sec.~\ref{sec:conc}.

\section{Review: Gauss-Bonnet inflation}\label{sec:GBI}
We consider the following action that involves the Einstein-Hilbert term and the GB term coupled to a canonical scalar field $\phi$ through the coupling
function $\xi(\phi)$~\cite{Guo:2010jr, Satoh:2010ep, Koh:2014bka, Jiang:2013gza, Koh:2016abf},
\begin{align}
S = &\int d^4x\sqrt{-g}
\left[\frac{1}{2\kappa^2} R - \frac{1}{2}g^{\mu\nu}
\partial_{\mu}\phi \partial_{\nu} \phi
 - V(\phi)-\frac12\xi(\phi) R_{\rm GB}^2\right],
\label{action}
\end{align}
where $R^{2}_{\rm GB} = R_{\mu\nu\rho\sigma} R^{\mu\nu\rho\sigma}- 4 R_{\mu\nu} R^{\mu\nu} + R^2$ is known as the GB term and
$\kappa^2 = 8\pi G=M_{\text{pl}}^{-2}$ is the reduced Planck mass. In the flat Friedmann-Robertson-Walker(FRW) universe with the scale factor $a$,
\begin{align}
ds^2 = - dt^2 + a^2\left(dr^2 +r^2 d\Omega^2\right),
\end{align}
the background dynamics of this system yields the Einstein and the field equations,
\bea \label{beq2a}
&& H^2 = \frac{\kappa^2}{3} \left(\frac{1}{2}\dot{\phi}^2 + V + 12\dot{\xi}H^3 \right)\,,\\
\label{beq3a}
&& \dot{H} = -\frac{\kappa^2}{2}\left[\dot{\phi}^2-4\ddot{\xi}H^2-4\dot{\xi}H\left(2\dot{H}-H^2\right) \right]\,, \\
&& \ddot{\phi} + 3 H \dot{\phi} + V_{\phi} +12 \xi_{\phi} H^2 \left(\dot{H}+H^2\right) = 0\,, \label{beq4a}
\ena
where the dot represents the derivative with respect to the cosmic time $t$,
$H \equiv \dot{a}/a$ denotes the Hubble parameter, $V_{\phi} = \partial V/\partial \phi,
\,\, \xi_{\phi} = \partial \xi/\partial \phi$, and $\dot{\xi}$ implies $\dot{\xi} = \xi_{\phi} \dot{\phi}$. The coupling function $\xi(\phi)$ is necessary to be a function of the scalar field; otherwise, the background dynamics will not be affected by the GB term.

In the context of slow-roll inflation, in which the friction term in Eq.~(\ref{beq4a}) is dominating and $\phi$ is considered to be slowly rolling down to the minimum of its potential, we define the slow-roll parameters 
\bea\label{eq:sl_param}
\epsilon\equiv-\frac{\dot{H}}{H^2}\,,\quad \eta\equiv\frac{\ddot{H}}{H\dot{H}}\,,\quad\zeta\equiv\frac{\dddot{H}}{H^2\dot{H}}\,,\quad\delta_1\equiv4\kappa^2\dot{\xi}H\,,
\quad\delta_2\equiv\frac{\ddot{\xi}}{\dot{\xi}H}\,,\quad\delta_3=\frac{\dddot{\xi}}{\dot{\xi}H^2}\,.
\ena
These parameters can be also expressed in terms of the potential and the coupling functions as
\bea\label{eq:srpote}
\epsilon &=& \frac{1}{2\kappa^2} \frac{V_{\phi}}{V}Q\,,\\
\eta &=& -\frac{Q}{\kappa^2}\left(\frac{V_{\phi\phi}}{V_{\phi}} + \frac{Q_{\phi}}{Q}\right)\,,\label{eq:srpotet}\\
\zeta&=& \frac{Q^2}{\kappa^4}\left[\left(\frac{V_{\phi\phi\phi}}{V_\phi} + \frac{V_{\phi\phi}}{2V}\right)+\left(\frac{3V_{\phi\phi}}{V_\phi}+\frac{V_\phi}{2V}\right)\frac{Q_{\phi}}{Q}+\frac{Q_\phi^2}{Q^2}+\frac{Q_{\phi\phi}}{Q}\right]\,,\\
\delta_1 &=& -\frac{4\kappa^2}{3} \xi_{\phi} V Q\,,\label{eq:sl_d1phi}\\
\delta_2 &=& -\frac{Q}{\kappa^2}\left(\frac{\xi_{\phi\phi}}{\xi_{\phi}}+\frac{1}{2}\frac{V_{\phi}}{V}+\frac{Q_{\phi}}{Q}\right)\,,\\
\delta_3 &=& \frac{Q^2}{\kappa^4}\left[\left(\frac{\xi_{\phi\phi\phi}}{\xi_\phi}+\frac{3\xi_{\phi\phi}V_\phi}{2\xi_\phi V}+\frac{V_{\phi\phi}}{2V}\right)+\left(\frac{3\xi_{\phi\phi}}{\xi_\phi}+\frac{2V_\phi}{V}\right)\frac{Q_\phi}{Q}+\frac{Q^2_\phi}{Q^2}+\frac{Q_{\phi\phi}}{Q}\right]
\,,\label{eq:sl_d2phi} 
\ena
where
\bea\label{eq:QofN}
Q &\equiv& \frac{V_{\phi}}{V} + \frac{4}{3}\kappa^4 \xi_{\phi} V\,.
\ena
The amount of the inflationary expansion is encoded in the number of $e$-folds,  $N$, 
\bea
\label{eq:ne}
N = \int^{t_\text{end}}_{t_\ast} Hdt \simeq \int^{\phi_\ast}_{\phi_\text{end}} \frac{\kappa^2}{Q} d\phi\,,
\ena
where the subscript ``$\ast$'' indicates the moment when a mode $k$  crosses the horizon during inflation.
The primordial power spectra of the scalar and the tensor perturbations at the time of horizon crossing are calculated in Ref.~\cite{Koh:2014bka} as
\bea
&&\mathcal{P}_S \simeq \frac{\csc^2 \nu_S \pi}{\pi z_S^2
\Gamma^2(1-\nu_S)} \frac{ a^2}{c_S^3 |\tau|^2 }
\biggl(\frac{c_S k|\tau|}{2}\biggr)^{3-2\nu_S},
\label{psfs}
\\
&&\mathcal{P}_T \simeq 8\frac{\csc^2 \nu_T \pi}{\pi z_T^2
\Gamma^2(1-\nu_T)} \frac{a^2}{c_T^3 |\tau|^2 }
\biggl(\frac{c_T k|\tau|}{2}\biggr)^{3-2\nu_T}\,,
\label{psft}
\ena
respectively, where $\tau$ is a conformal time, which is related to the cosmic time via $\tau=\int a^{-1} dt$. The quantities $\nu_A$, $c_A$, and $z_A$ with $A=\{S, T\}$ are given by
\bea
&& \nu_S \simeq \frac{3}{2} + \epsilon +
\frac{2\epsilon (2\epsilon +\eta) - \delta_1 (\delta_2 -\epsilon)}
{4\epsilon-2\delta_1}\,, \quad
\qquad \nu_T \simeq \frac{3}{2} +\epsilon\,,\\
&&c_S^2 = 1 - \frac{ (4 \epsilon
+ \delta_1 (1- 4\epsilon -\delta_2) )\Delta^2}{4\epsilon - 2\delta_1
-2\delta_1 (2\epsilon -\delta_2) +3 \delta_1 \Delta}, \quad
\, \, \,c_T^2 = 1 + \frac{\delta_1 (1 - \delta_2)}{1-\delta_1},\label{eq:proSpd}
\\
&&z_S = \sqrt{\frac{a^2}{\kappa^2}
\frac{2\epsilon -\delta_1(1+2\epsilon-\delta_2)
+\frac{3}{2}\delta_1 \Delta}{(1-\frac{1}{2}\Delta)^2}},
\quad\,\,\quad z_T = \sqrt{\frac{a^2}{\kappa^2}(1-\delta_1)},\label{eq:zA}
\ena
where $\Delta=\delta_1/(1-\delta_1)$. The observable quantities such as the spectral indices of the scalar and the tensor perturbations, their running spectral indices, and the  tensor-to-scalar ratio are derived, respectively, as follows:
\bea\label{eq:nsntr}
&&n_S-1\simeq-2\epsilon-\frac{2\epsilon(2\epsilon+\eta)-\delta_1(\delta_2-\epsilon)}{2\epsilon-\delta_1}\,,\,\,\,\quad  n_T\simeq -2\epsilon\,,\nonumber\\
&&\alpha_S=-2\epsilon(2\epsilon+\eta)+\left[\frac{2\epsilon(2\epsilon+\eta)-\delta_1(\delta_2-\epsilon)}{2\epsilon-\delta_1}\right]^2-\frac{2\epsilon(8\epsilon^2+7\epsilon\eta+\zeta)+\delta_1(\epsilon^2+\epsilon\eta+\epsilon\delta_2-\delta_3)}{2\epsilon-\delta_1}\,,\nonumber \\
&&\alpha_T=-2\epsilon(2\epsilon+\eta)\,,\,\,\, \quad r\simeq8(2\epsilon-\delta_1)\,,
\ena
where $n_S-1 = d\ln\mathcal{P}_S/d\ln k,\,\, n_T = d\ln\mathcal{P}_T/d\ln k,\,\, \alpha_S = d n_S/d\ln k,\,\, \alpha_T = d n_T/d\ln k$, and $r=\mathcal{P}_T/\mathcal{P}_S$.
If the potentials $V(\phi)$ and the coupling function $\xi(\phi)$ are given, it is straightforward to calculate Eq.~(\ref{eq:nsntr}) by using Eqs.~(\ref{eq:srpote})---(\ref{eq:sl_d2phi}). Thus, the theoretical predictions of any particular model of inflation obtained through Eq.~(\ref{eq:nsntr}) can be tested by the observational data~\cite{Planck:2013jfk, Ade:2015lrj}.

\section{Gauss-Bonnet inflation models}\label{sec:class}
The standard single-field models of slow-roll inflation with a canonical kinetic term, as discussed in Refs.~\cite{Planck:2013jfk, Ade:2015lrj}, predict a slightly red-tilted primordial tensor power spectrum, i.e. $n_T<0$ with $|n_T|\ll1$. However, the spectrum of the inflationary tensor perturbations could have a blue tilt $n_T>0$~\cite{Ade:2015lrj, Brandenberger:2014faa, Hwang:2005hb}. Therefore, any evidence of the blue-tilted tensor mode spectrum would support nonstandard models of inflation. In this section, we consider two types of  inflation model with a GB term based on their predictions for the $n_T$, a positive and a negative.\footnote{We exclude the scale-invariant case where $n_T=0$ in the present work.} The models that predict the inflationary tensor power spectrum with a red tilt ($n_T < 0$) are classified as the \emph{model I}, whereas those that predict the blue-tilted inflationary tensor power spectrum are grouped as the \emph{model II}. In order for the tensor mode spectrum to have a red tilt (blue tilt), the slow-roll parameter $\epsilon$ in Eq.~(\ref{eq:nsntr}) has to be negative (positive). $\epsilon$ could be positive if the potentials and the coupling functions must satisfy the following conditions from Eq.~(\ref{eq:srpote}):

\[\left\{
\begin{array}{c}
\xi_\phi > -\frac{3}{4\kappa^4}\frac{V_\phi}{V^2} \quad \text{for} \quad V_\phi >0, \\
\,\,\xi_\phi < -\frac{3}{4\kappa^4}\frac{V_\phi}{V^2} \quad \text{for} \quad V_\phi <0\,.\nonumber
\end{array}
\right. \]
Among several successful inflationary models that satisfy these conditions~\cite{Koh:2014bka, Jiang:2013gza, Guo:2010jr}, we consider the power law potential with an  inverse monomial coupling and identify this model as model I. The inflaton potential and the coupling function for model I are given by
\bea
V(\phi) = \frac{V_0}{\kappa^4}(\kappa \phi)^n\,, \quad \xi(\phi)=\xi_0 (\kappa\phi)^{-n}\,,\label{eq:potandInvGB}
\ena
respectively, where {\color{black}$V_0$ is a dimensionless constant and $n>0$ is assumed}.
From Eqs.~(\ref{eq:ne}) and (\ref{eq:nsntr}), the observable quantities are obtained in terms of $N_\ast$ as
\bea\label{eq:obsofN}
&&n_S -1 = -\frac{2(n+2)}{4N_\ast+n}\,,\,\,\, n_T = -\frac{2n}{4N_\ast+n}\,,\,\,\, r=\frac{16n(1-\alpha)}{4N_\ast+n},\\
&& \alpha_S = -\frac{8(n+2)}{(4N_\ast+n)^2}\,,\,\,\, \alpha_T = -\frac{8n}{(4N_\ast+n)^2}\,, \nonumber
\ena
where $\alpha\equiv4V_0\xi_0/3$. One can see from the above equations that the tensor spectral index for model I is always negative as long as $n>0$; hence, the inflationary tensor  power spectrum has a red tilt, $n_T<0$. The tensor-to-scalar ratio $r$ is suppressed for a positive $\alpha$, while it is enhanced for a negative $\alpha$.

In order for $\epsilon$ to be negative, the potential and the coupling functions satisfy the following conditions from Eq.~(\ref{eq:srpote}):
\[\left\{\begin{array}{c}
\xi_\phi < -\frac{3}{4\kappa^4}\frac{V_\phi}{V^2} \quad \text{for} \quad V_\phi >0 ,\\
\,\,\xi_\phi > -\frac{3}{4\kappa^4}\frac{V_\phi}{V^2} \quad \text{for} \quad V_\phi <0\,.\nonumber
\end{array}
\right. \]

This kind of model is identified as model II. We take the following potential and the coupling function for this type, which was first introduced in Ref.~\cite{Koh:2016abf}, as~\footnote{The shape of the potential in Eq.~(\ref{eq:RecForm}) is similar to that of the "T-model" in Ref.~\cite{Kallosh:2013hoa} in the $\mu\rightarrow0$ limit. For $\mu\neq0$, there appears a small bump on the side of the potential, which differs the model from the "T-model".}
\bea\label{eq:RecForm}
V(\phi)=\frac{1}{\kappa^4}\left[\text{tanh}\left(\kappa\phi\right)+\sqrt{\mu}\,\text{sech}\left(\kappa\phi\right)\right]^2\,, \quad 
\xi(\phi)=\frac{3\left[\text{sinh}^2(\kappa\phi)-\frac{1}{\sqrt{\mu}}\text{sinh}(\kappa \phi)\right]}{4\left[\sqrt{\mu}+\text{sinh}\left(\kappa\phi\right)\right]^2}\,,
\ena
where $\mu>0$ is assumed.
From Eqs.~(\ref{eq:ne}) and (\ref{eq:nsntr}), the observable quantities are obtained as,
\bea\label{eq:obsofNm4}
&&n_S -1 =- \frac{2}{N_\ast+\mu}\,,\,\,\, n_T = \frac{2\mu(N_\ast-1)}{(N_\ast+\mu)(N_\ast^2+\mu)}\,,\,\,\, r=\frac{8}{N_\ast^2+\mu}\,,\\
&&  \alpha_S = -\frac{2}{(N_\ast+\mu)^2}\,,\,\,\, \alpha_T = -\frac{2}{(N_\ast+\mu)^2}+\frac{2(N_\ast^2-\mu)}{(N_\ast^2+\mu)^2}\,. \nonumber
\ena
Since $n_S$ is well constrained by the current observation, the range of the model parameter $\mu$ can be determined from Eq.~(\ref{eq:obsofNm4}) to be $\mu=2/(1-n_S)-N_\ast\sim\mathcal{O}(10)$ where $40\leq N_\ast\leq70$ is assumed. $n_T$ is positive as long as $N_\ast>1$, which is necessary condition for inflation to successfully solve the horizon and the flatness problems of standard big bang cosmology. Thus, the inflationary tensor power spectrum of model-II always has a blue tilt, $n_T>0$.

\begin{figure}[H]
\centering
\includegraphics[width=0.5\textwidth]{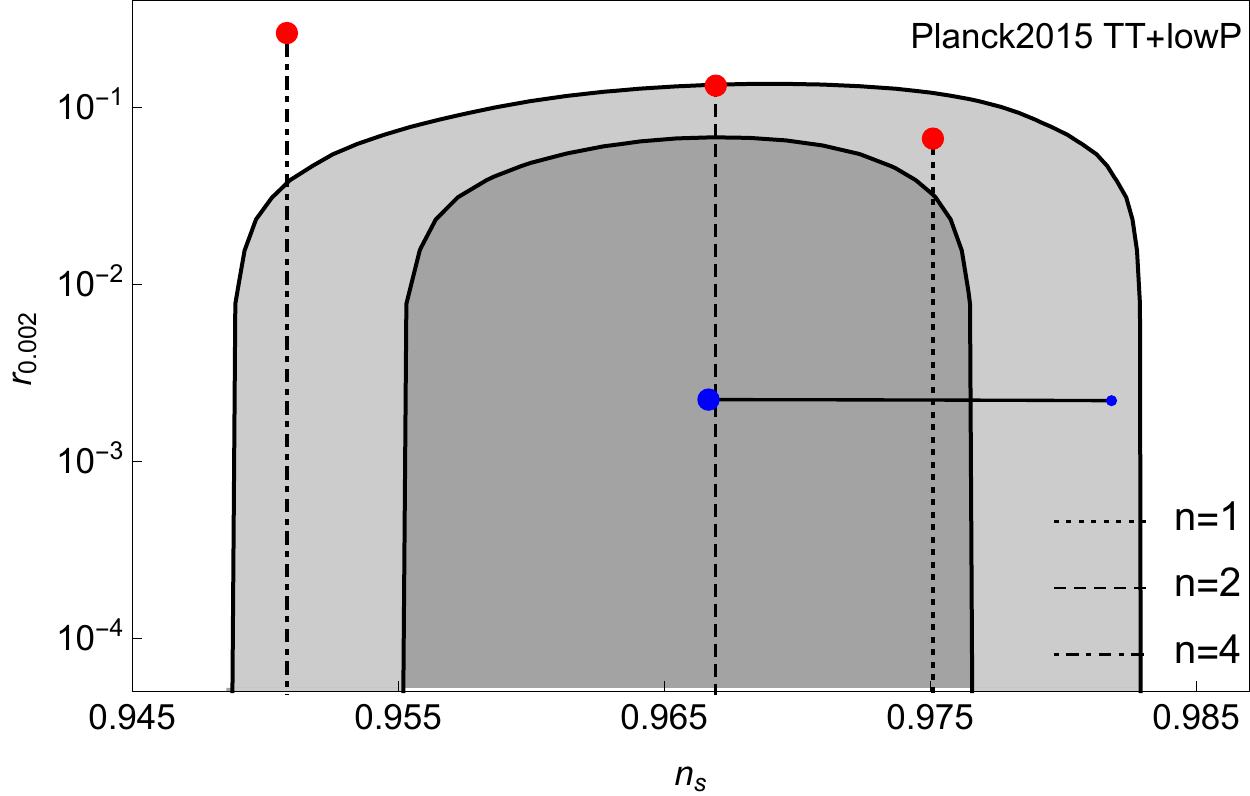}
\caption{The marginalized $68\%$ and $95\%$ confidence level contours for $n_S$ and $r_{0.002}$ from Planck2015 TT+lowP~\cite{Ade:2015lrj} and the theoretical predictions of models in Eqs.~(\ref{eq:potandInvGB}) and (\ref{eq:RecForm}). The red dots indicate model I with fixed $\alpha=0$ but varying $n$. The parameter $\alpha$ then grows from zero to unity along each $n=1~\text{(dotted)}$, $n=2~\text{(dashed)}$, and $n=4~\text{(dot-dashed)}$ line. For model II, the black solid line, the values of $\mu$ also increase from the larger blue end point to a smaller one, where $10^{-4}<\mu\leq50$. The $e$-folding number is set to $N_\ast=60$ along each line.}\label{fig:compared}
\end{figure}

We plot the theoretical predictions of model I for $n=1\,\,(\text{dotted line}),\,\,2\,\,(\text{dashed line})$, and $n=4\,\,(\text{dot-dashed line})$ and model II (solid line) when $N_{*}=60$ against the observational data~\cite{Ade:2015lrj} in the $n_S-r$ plane in Fig.~\ref{fig:compared}. Red dots indicate the predictions for $\alpha =0$, which corresponds to the standard single-field slow-roll inflation model. One can find from Eq.~(\ref{eq:obsofN}) that $n_s$ is independent of $\alpha$ but $r$ is decreasing as $\alpha\,\, (> 0)$ is increasing and $r=0$ if $\alpha =1$. Therefore, the vertical dotted, dashed, and dot-dashed lines in Fig.~\ref{fig:compared}  represent the effect of $\alpha$ on $r$ for $n=1,\,\,2$, and $4$, respectively. In this work, we limit our interests to only $ 0\leq \alpha \leq 1$. The predictions of model II are plotted in solid line with $0<\mu\leq50$, where the big blue dot corresponds to $\mu=0$. By using the marginalized mean value for $n_S=0.9655\pm0.0062$ from the observation~\cite{Ade:2015lrj}, we get the upper limit of the tensor-to-scalar ratio for model II to be $r\simeq0.0032$ for $N_\ast=50$ and $r\simeq0.0022$ for $N_\ast=60$ which are nearly insensitive on $\mu$. Further details of the each inflation model can be found in the corresponding references~\cite{Koh:2014bka, Jiang:2013gza, Koh:2016abf, Guo:2010jr}.

\section{Primordial Gravitational Waves induced by the blue-tilted and red-tilted tensor spectra}\label{sec:GWES}
We discussed two types of GB inflation model in the previous section. In this section, we calculate the energy spectrum of the PGW background for selected models: model I and model II. We start the present section by reviewing a formalism to calculate the energy spectrum of the PGW background. The PGWs are described by a tensor part of the metric fluctuations in the linearized flat FRW metric of the form
\bea\label{FRWmetric}
ds^2=a^2(\tau)\left[-d\tau^2+(\delta_{ij}+h_{ij})dx^idx^j\right],
\ena
where $h_{ij}$ is symmetric under the exchange of indices and satisfies the transverse-traceless condition $\partial_ih^{ij}=0\,,\,\delta^{ij}h_{ij}=0$. The tensor perturbation can be expanded in Fourier space as
\bea
h_{ij}(\tau,\mathbf{x})&=& \sum_{\lambda}\int\frac{d\,\mathbf{k}}{(2\pi)^{3/2}}\epsilon^{\lambda}_{ij} h_{\lambda, \mathbf{k}}(\tau) e^{i\mathbf{k}\mathbf{x}}\,,
\ena
where $\lambda$ denotes each polarization state of the tensor perturbations and $\epsilon_{ij}^{\lambda}$ is the symmetric polarization tensor, which satisfies the transverse-traceless condition and is normalized by the relation $\sum_{i,j}\epsilon_{ij}^{\lambda}\left(\epsilon_{ij}^{\lambda'}\right)^*=2\delta^{\l\l '}$. The GW energy density $\rho_{GW}$ is defined by $\rho_{GW}=-T^0\,_0$ and can be written as
\bea\label{rhoGW}
\rho_{GW}=\frac{M_p^2}{4}\int d\ln k \left(\frac{k}{a}\right)^2 \frac{k^3}{\pi^2}\sum_{\l}\langle h_{\l,\,k}^\dagger h_{\l,\,k}\rangle .
\ena
where the bracket $\langle\cdots\rangle$ indicates the spatial average. The strength of GW is characterized by their energy spectrum, which is written by
\bea\label{gwEs}
\O_{GW}(k)=\frac{1}{\rho_{\text{crit}}}\frac{d\rho_{GW}}{d\ln k},
\ena
where $\rho_{\text{crit}}=3H_0^2M_p^2$ is the critical density and $H_0$ is the present Hubble constant, which is measured by the observation as $H_0=100 h_0\,\text{km}\,\text{s}^{-1}\text{Mpc}^{-1}$ with $h_0=0.6731$~\cite{Ade:2015lrj}.
By using Eqs. (\ref{rhoGW}) and~(\ref{gwEs}), we rewrite
\bea\label{gwEsofk1}
\O_{GW}(k)=\frac{k^2}{12H_0^2}P_{T}(k),
\ena
where $P_{T}$ is the power spectrum of the PGW observed today and is related to that of the inflationary one $\mathcal{P}_{T}(k)$ through the transfer function $\mathcal{T}(k)$ as follows
\bea\label{eq:tpowspec0}
P_{T}\equiv\frac{k^3}{\pi^2}\sum_{\l}\langle h_{\l,\,k}^\dagger h_{\l,\,k}\rangle=\mathcal{T}^2(k)\mathcal{P}_{T}(k).
\ena
The inflationary power spectrum for the tensor perturbations can be parameterized as follows:
\bea\label{eq:tpowspec}
\mathcal{P}_T(k)=\mathcal{P}_{T}(k_\ast)\left(\frac{k}{k_{\ast}}\right)^{n_T+\frac{\alpha_T}{2}\ln (k/k_{\ast})},
\ena
where $k_{\ast}$ is the reference pivot scale.
The amplitude $\mathcal{P}_{T}(k_\ast)$ is often characterized by the tensor-to-scalar ratio $r$ as $\mathcal{P}_T(k_\ast)=r\mathcal{P}_S(k_\ast)$, where $\mathcal{P}_S(k_\ast)$ is well measured by the observation as $\ln(10^{10}\mathcal{P}_S)=3.089^{+0.024}_{-0.027}$ at $k_\ast=0.05\,\,\text{Mpc}^{-1}$~\cite{Ade:2015lrj}.

The transfer function reflects the evolution of GWs after horizon reentry; hence, it depends on the thermal history of the Universe. One can attempt a task to calculate the transfer function by numerically integrating the evolution equation for  the PGW following Refs.~\cite{Turner:1993vb, Zhao:2006mm, Watanabe:2006qe, Nakayama:2008wy, Kuroyanagi:2008ye, Giovannini:2008tm, Nakayama:2009ce, Kuroyanagi:2014nba, Tumurtushaa:2016ars, Koh:2015brl, Kuroyanagi:2011fy, Buchmuller:2013lra}. The evolution equation of the PGW for our models is given by~\cite{Koh:2016abf}
\bea\label{waveeq1}
h_{\l,k}''+2\frac{z_T'}{z_T} h_{\l,k}'+k^2 c_T^2 h_{\l,k}=0,
\ena
where $'\equiv d/d\t$. The mode solutions to this equation have qualitative behavior in two regimes~\cite{Hwang:2005hb}: either outside the horizon $(k\ll aH)$ where the amplitude of $h_{\l,k}$ remains constant,
\begin{align}
h_{\lambda,k} = C(k),
\end{align}
 or inside the horizon $(k\gg aH)$ where the amplitude begins to damp
\begin{align}
h_{\lambda,k} = \frac{1}{z_T} [c_1 e^{ic_T k\tau} + c_2 e^{-ic_T k\tau}].
\end{align}
The exact solutions for $z \propto |\tau|^q,\,\, c_T^2 = \text{const}$ are
\begin{align}
h_{\lambda,k} = \frac{\sqrt{\pi |\tau|}}{2z_T} [ c_1(k) H_{\nu}^{(1)} (c_T k|\tau|)
+ c_2(k) H_{\nu}^{(2)} (c_T k|\tau|)],
\end{align}
where $\nu= 1/2 -q$.

\begin{figure}[H]
\centering
\includegraphics[width=0.5\textwidth]{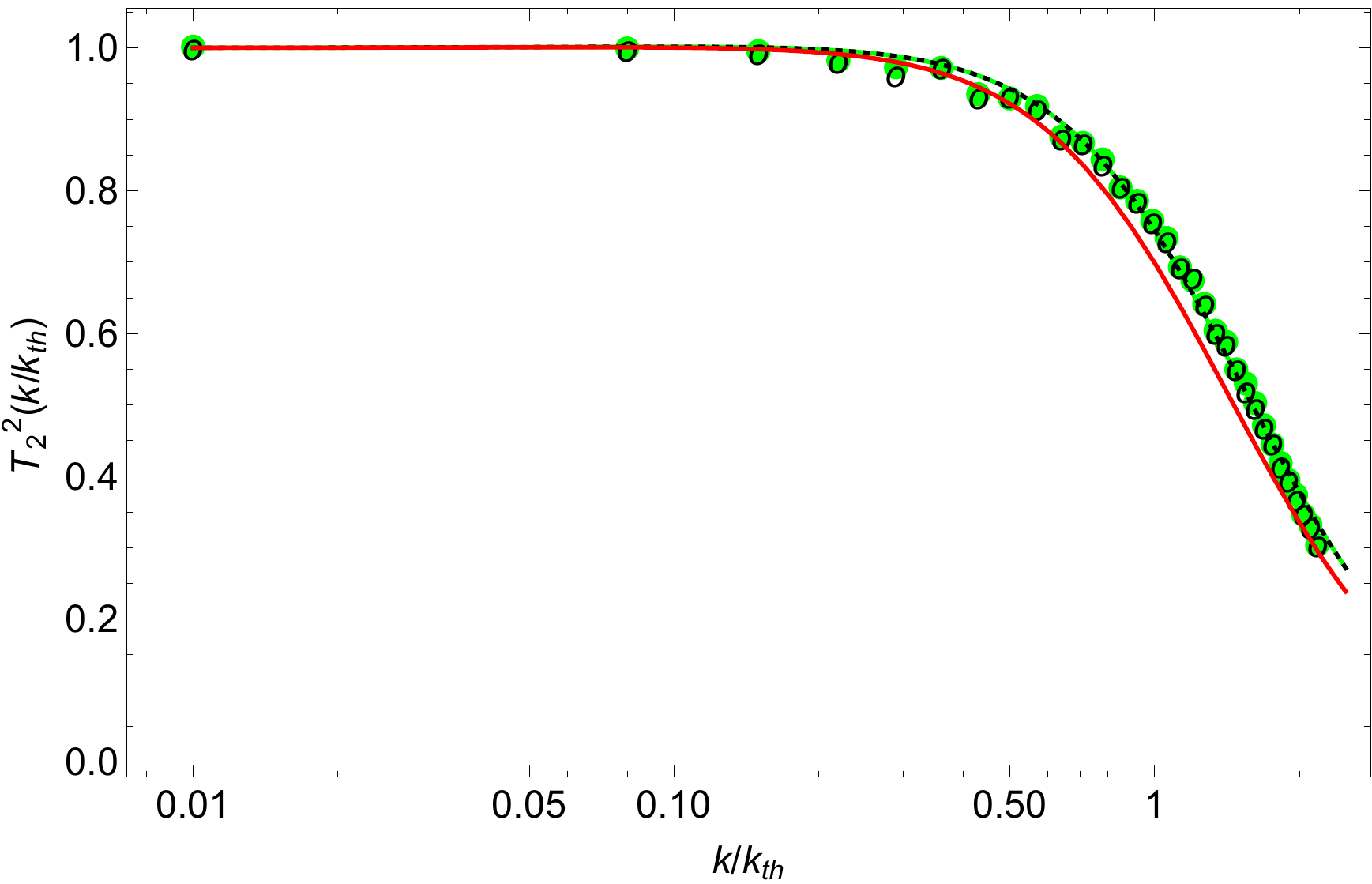}
\caption{Transfer functions given in Eq.~(\ref{eq:transF2}) for model I and model II in comparison with that of Ref.\cite{Kuroyanagi:2014nba}.}\label{fig:TransFunc}
\end{figure}

\begin{figure}[H]
\centering
\subfigure[{\tiny \emph{Model-I} with varying model parameter $\alpha$.}]
{\includegraphics[width=0.48\textwidth]{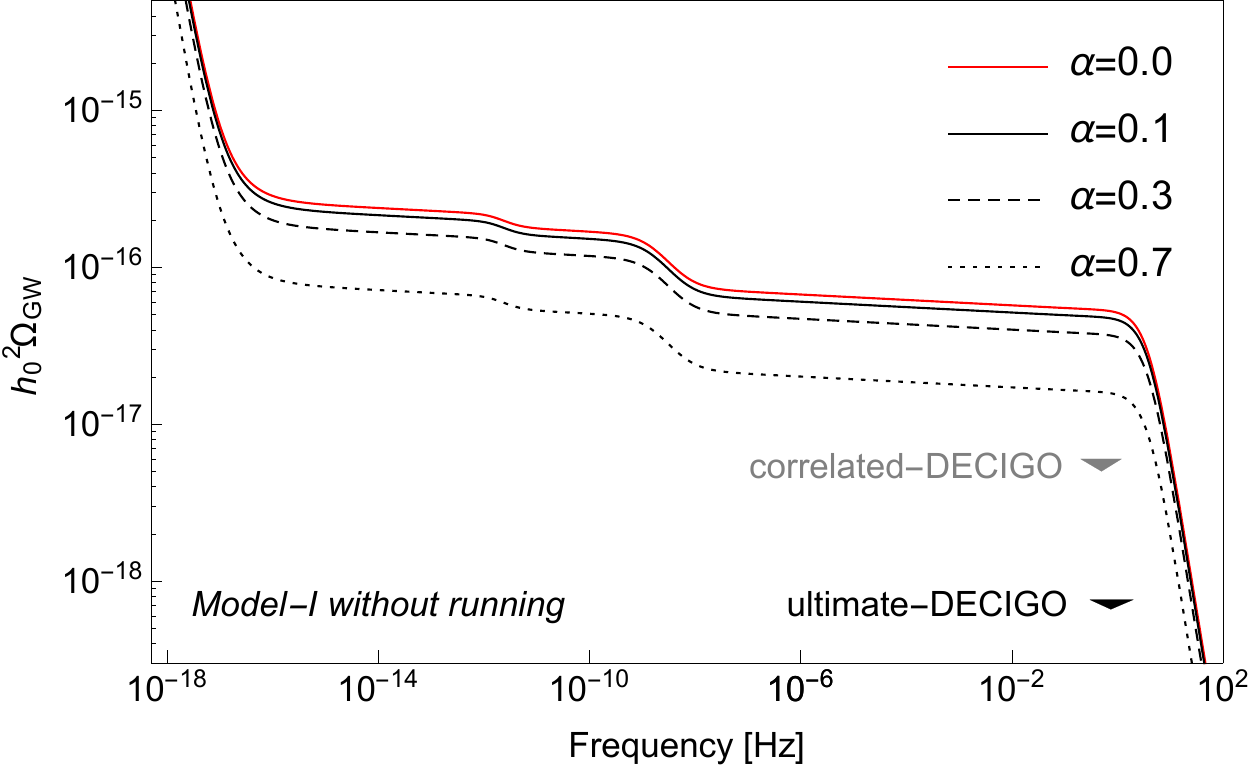}\label{fig:fig2aa}}
\subfigure[{\tiny \emph{Model-II} with varying model parameter $\mu$.}]
{\includegraphics[width=0.48\textwidth]{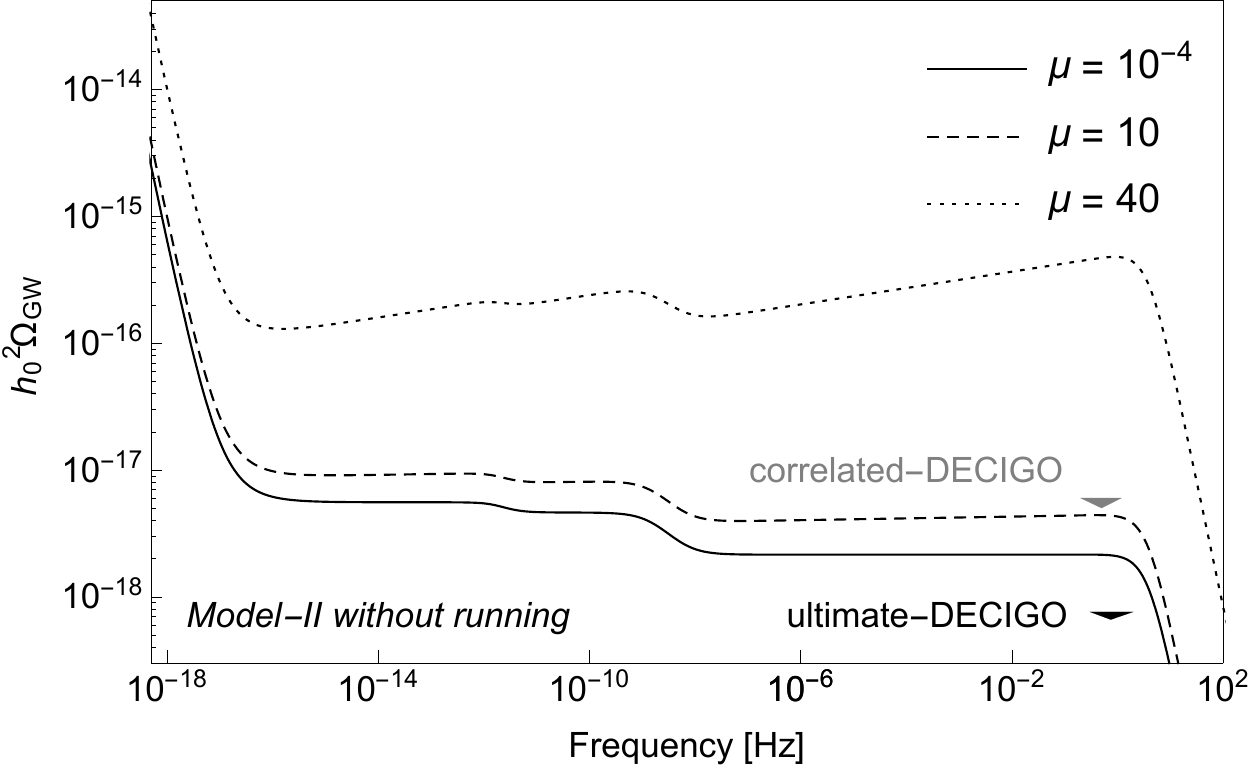}\label{fig:fig2bb}}\\
\subfigure[{\tiny \emph{Model-I} with $\alpha_T=0$ and $\alpha_T\simeq-0.0003$.}]
{\includegraphics[width=0.48\textwidth]{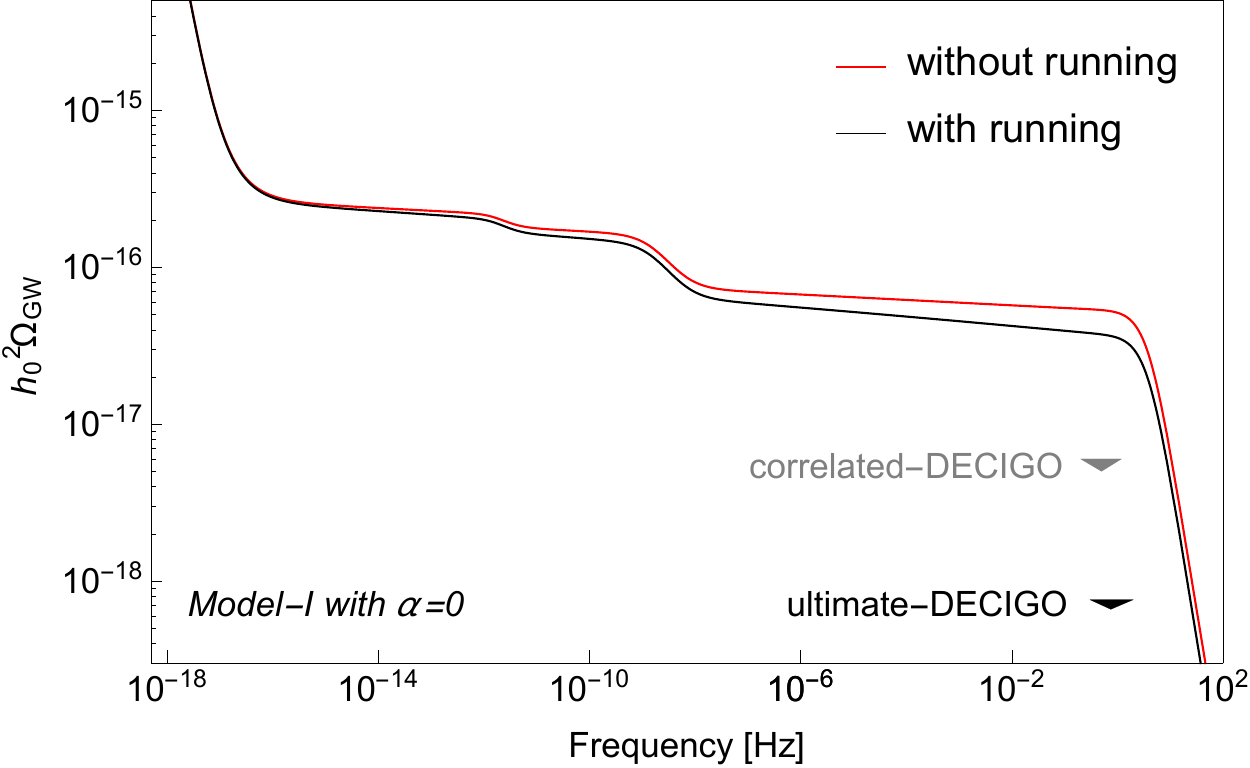}\label{fig:fig2c}}
\subfigure[{\tiny \emph{Model-II} with $\alpha_T=0$ and $\alpha_T\simeq0.0037$.}]
{\includegraphics[width=0.48\textwidth]{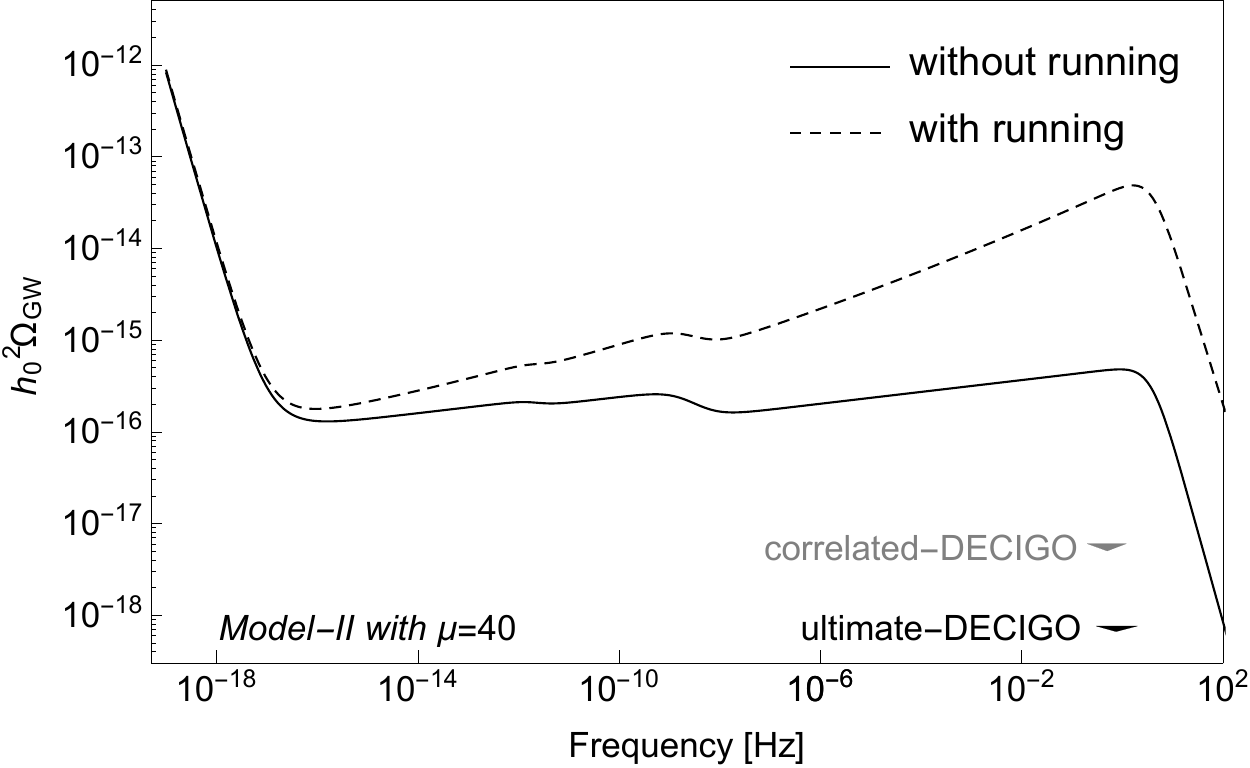}\label{fig:fig2d}}\\
\subfigure[{\tiny \emph{Model-I} with $\alpha=0$ and varying $T_\text{th}$.}]
{\includegraphics[width=0.48\textwidth]{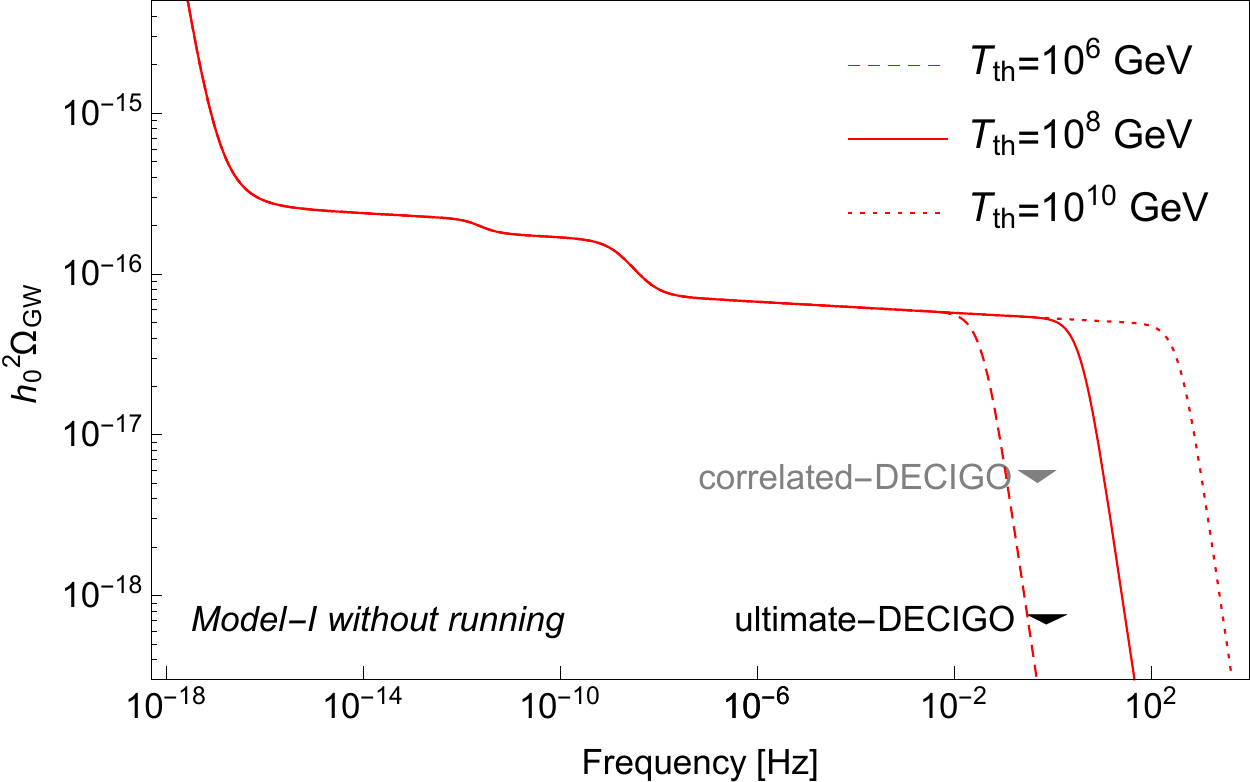}\label{fig:fig2e}}
\subfigure[{\tiny \emph{Model-II} with $\mu=40$ and varying $T_\text{th}$.}]
{\includegraphics[width=0.48\textwidth]{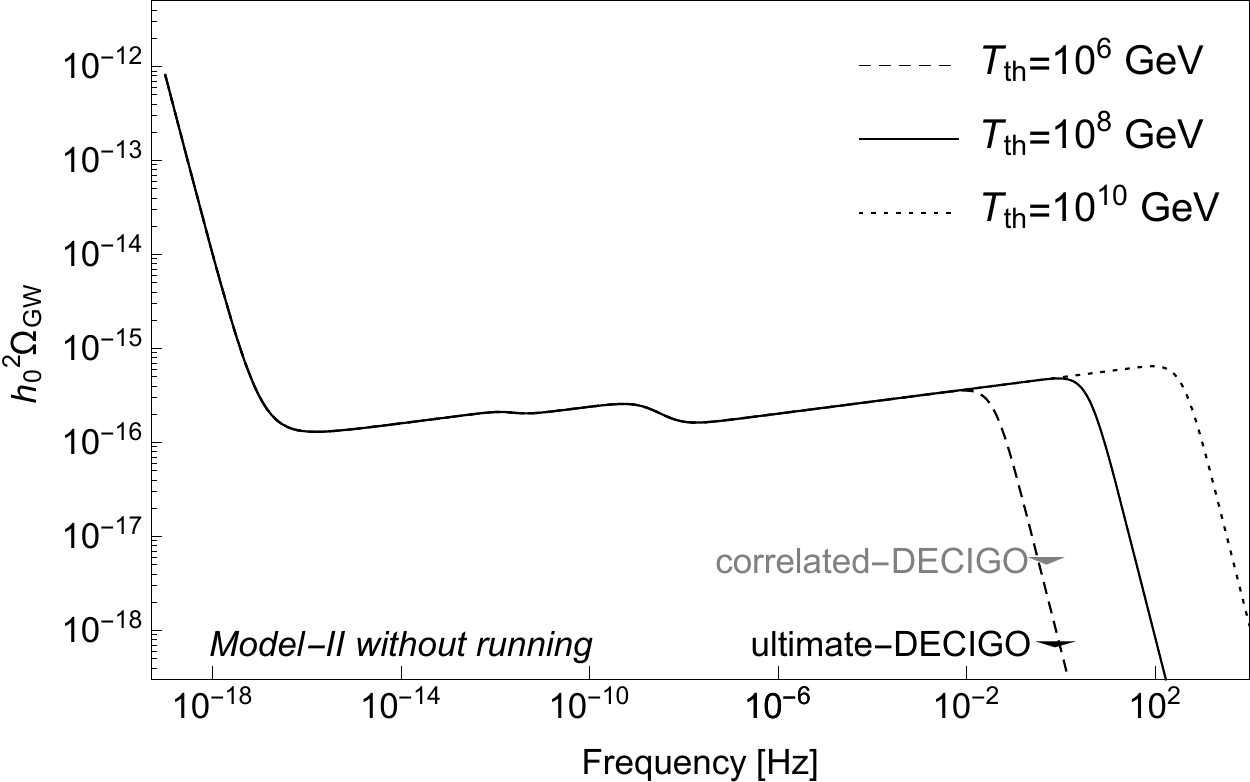}\label{fig:fig2f}}
\caption{The frequency dependence of the energy spectrum for model I and model II together with sensitivities of DECIGO~\cite{Kudoh:2005as}. We set $n_S=0.9655$ and $T_\text{th}=10^{8}$ GeV for (a)--(d). The red lines in (a), (c), and (e) indicate the absence of the GB term.}\label{fig:GWconst}
\end{figure}

For modes that reenter the horizon during the matter-dominated (MD) era, the solution to Eq.~(\ref{waveeq1}) evolves as $h_{\lambda,k} \sim 3j_1(k\tau)/(k\tau)$~\cite{Turner:1993vb, Zhao:2006mm, Watanabe:2006qe, Hwang:2005hb} {\color{black} for $z_T \sim a \sim \tau^2$, where we have assumed the GB effect is negligible during the matter-dominated era}.  The changes in the relativistic degrees of freedom $g_\ast(T_\text{in})$ and their counterpart $g_{\ast s}(T_\text{in})$ for entropy give another damping factor; see the third and fourth terms on the right-hand side of Eq.~(\ref{waveeq2})~\cite{Watanabe:2006qe}. Here $T_\text{in}$ is the temperature of the Universe at which the mode reenters the horizon. The amplitude of modes that reenter the horizon before matter and radiation equality would be suppressed by the expansion of the Universe. The suppression should be larger for modes that reenter the horizon earlier, as $g_\ast$ and  $g_{\ast s}$ would be larger than those for modes that reenter the horizon later; see the mid-frequency range in Fig.~\ref{fig:GWconst}. However, the modes that reenter the horizon during the MD era should not be affected by changes in the $g_\ast$ and $g_{\ast s}$, as they do not change during the MD era~\cite{Watanabe:2006qe}. Taking all these into account, a good fit to the transfer function is given by
\bea\label{waveeq2}
\mathcal{T}^2(k)=\O_m^2\left(\frac{3j_1(k\tau_0)}{k\tau_0}\right)^2\left(\frac{g_{\ast}(T_\text{in})}{g_{\ast 0}}\right) \left(\frac{g_{\ast s0}}{g_{\ast s}(T_\text{in})}\right)^{4/3}\mathcal{T}_1^2\left(\frac{k}{k_\text{eq}}\right)\mathcal{T}_2^2\left(\frac{k}{k_\text{th}}\right)\,,
\ena
where $\O_mh_0^2=0.1344$ is the matter density of the Universe, $g_{\ast s}(T_{\text{th}})$ is the effective number of light species for the entropy at the end of reheating and $T_{\text{th}}$ is the reheating temperature, and the subscript ``0'' denotes that the quantity is evaluated at the present time~\cite{Ade:2015lrj}. In the $k\tau_0\rightarrow0$ limit, the first spherical Bessel function becomes $j_1(k\tau_0)=1/(\sqrt{2}k\tau_0)$, where $\tau_0\simeq2H_0^{-1}$ is the present conformal time. The transfer functions $\mathcal{T}_1^2(k/k_\text{eq})$ and $\mathcal{T}_2^2(k/k_\text{th})$ are calculated by numerically integrating Eq.~(\ref{waveeq1}). For modes that reenter the horizon before or after matter and radiation equality, we get
\bea\label{eq:transF1}
\mathcal{T}_1^2\left(\frac{k}{k_\text{eq}}\right)&=&1+1.65\left(\frac{k}{k_\text{eq}}\right)+1.92\left(\frac{k}{k_\text{eq}}\right)^2\,,
\ena
where $k_\text{eq}=7.3\times 10^{-2} \Omega_m h_0^2\,\,\text{Mpc}^{-1}$ is the comoving wave numbers corresponding to the scale at the time of matter and radiation equality. The transfer function for modes that reenter the horizon after the end of inflation and before the end of reheating is calculated as
\bea\label{eq:transF2}
\mathcal{T}_2^2\left(\frac{k}{k_\text{th}}\right)&=&\left[1+\gamma \left(\frac{k}{k_\text{th}}\right)^{\frac{3}{2}}+\sigma\left(\frac{k}{k_\text{th}}\right)^2\right]^{-1}\,,
\ena
where $k_\text{th}=1.7\times10^{13}\,\,\text{Mpc}^{-1}\left(g_{\ast s}(T_\text{th})/106.75\right)^{1/6}\left(T_\text{th}/10^6\,\,\text{GeV}\right)$ is the comoving wave number corresponding to the scale at the time of the completion of reheating when the Universe became radiation dominated.\footnote{The coefficients of the transfer functions depend on the choice of reference wave number, $k_\text{eq}$ and $k_\text{th}$.}
The coefficients $\gamma$ and $\sigma$ are different for different inflationary models. In Fig.~\ref{fig:TransFunc}, we plot the result of Eq.~(\ref{eq:transF2}) for model I (dashed black line) and model II (green line) in comparison with Eq.~(2.16) of Ref.~\cite{Kuroyanagi:2014nba} (red line), which is the case where the GB term is absent. In the figure, ``o'' and ``\tikzcircle{2pt}'' denote the numerical solutions of the transfer function, while the dashed black and green lines are the fitted transfer functions of Eq.~(\ref{eq:transF2}) for model I and model II, respectively. One can see that the transfer functions for model I and model II have the same shape with coefficients of $\gamma\simeq-0.23$ and $\sigma\simeq0.58$.

By substituting Eqs.~(\ref{eq:tpowspec}) and~(\ref{waveeq2}) into Eq.~(\ref{eq:tpowspec0}) and then into Eq.~(\ref{gwEsofk1}), we obtain
\bea\label{ESofGWofk}
h_0^2\O_{GW}=\frac{3h_0^2}{32\pi^2H_0^2\tau_0^4f^2}
\O_m
^2 \mathcal{T}_1^2\left(\frac{f}{f_\text{eq}}\right)\,\mathcal{T}_2^2\left(\frac{f}{f_\text{th}}\right) r \mathcal{P}_S\left(\frac{f}{f_{\ast}}\right)^{n_T+\frac{\alpha_T}{2}\ln (f/f_{\ast})}\,,
\ena
where the frequency relation $k=2\pi f$ is used. The quantities $n_T$, $\alpha_T$, and $r$ are the functions of the slow-roll parameters.  However, they can be expressed in terms of $n_S$ and the model parameters $\alpha$ and $\mu$ through Eqs.~(\ref{eq:obsofN}) and (\ref{eq:obsofNm4}) once the potential and coupling functions are given. For selected models from the last section, we obtain
\bea
\text{model I:} && r = - \frac{8n(1-\alpha)(n_S-1)}{n+2}\,,\, n_T = \frac{n(n_S-1)}{n+2}\,,\, \alpha_T=-2n\left(\frac{n_S-1}{n+2}\right)^2\,;\\
\text{model II:}  &&r = \frac{8(n_S-1)^2}{\mu(n_S-1)^2+[2+\mu(n_S-1)]^2}\,,\nonumber\\ &&n_T = \frac{\mu(n_S-1)^2[n_S+1+\mu(n_S-1)]}{4+\mu(n_S-1)[n_S+3+\mu(n_S-1)]}\,,\\&& \alpha_T=-\frac{(n_S-1)^2}{2}+2\left[\left(\frac{2+\mu(n_S-1)}{n_S-1}\right)^2-\mu\right]\left[\left(\frac{2+\mu(n_S-1)}{n_S-1}\right)^2+\mu\right]^{-2}\,.\nonumber
\ena

We plot the frequency dependence of the energy spectrum from Eq.~(\ref{ESofGWofk}) for model I with $n=2$ and model II in Fig.~\ref{fig:GWconst}. Along with the predicted energy spectrum, the sensitivities of the future space-based laser interferometric experiment DECIGO~\cite{Kudoh:2005as}, including correlated-DECIGO and ultimate-DECIGO (single), are presented.

As we can see in Fig.~\ref{fig:GWconst}, both model I and model II predict an observable GW spectrum around a frequency $0.1-10$ Hz. The amplitude of the PGW spectrum sourced by model I is suppressed as $\alpha$ increases; see Fig.~\ref{fig:fig2aa}. On the other hand, for model II, the amplitude is enhanced for increasing values of $\mu$ as is seen in Fig.~\ref{fig:fig2bb}. In the figure, we used observationally preferred values of $\alpha$ and $\mu$ from Fig.~\ref{fig:compared}, \emph{namely}, $0\leq\alpha\leq1$ for model I and $0<\mu\lesssim50$ for model II. The running of the spectral index of the tensor perturbations, $\alpha_T$, further suppresses (enhances) the spectrum for model I (model II) as the frequency increases as is seen in Fig.~\ref{fig:fig2c} [Fig.~\ref{fig:fig2d}], where the spectra with and without $\alpha_T$ are compared. This suppression (enhancement) implies that $\alpha_T$ is negative (positive).

Although the amount of suppression or enhancement of the energy spectrum is determined by the value of the model parameters ($\alpha$ and $\mu$), the bending frequency of the spectrum depends only on the reheating temperature $T_\text{th}$. In Fig.~\ref{fig:GWconst}, \emph{for simplicity}, we set the reheating temperature to be $T_\text{th}=10^{8}$ GeV {\color{black} for Figs. ~\ref{fig:fig2aa}--\ref{fig:fig2d}} ; hence, the spectrum is significantly suppressed near frequency $f_\text{th}\simeq2.6$ Hz, which resides in the most sensitive frequency range ($0.1$--$10$ Hz) of the planned future space-based experiment DECIGO. The lower limit of the reheating temperature by DECIGO is about $T_\text{th}\gtrsim10^{6}$ GeV [see Figs.~\ref{fig:fig2e} and \ref{fig:fig2f}]. We thus use this bound in our analysis of the reheating parameters including the temperature $T_\text{th}$, equation of state $\omega_\text{th}$ , and the duration of reheating $N_\text{th}$  in the next section. As we mentioned before, the suppression of the spectrum between $10^{-12}~\text{Hz}~\lesssim f\lesssim10^{-8}$ Hz in Fig.~\ref{fig:GWconst} is due to the changes in the relativistic degrees of freedom $g_\ast$ and its counterpart for entropy $g_{\ast s}$, the third and fourth terms in Eq.~(\ref{waveeq2})~\cite{Watanabe:2006qe}.

\section{Constraints on the reheating parameters from the primordial gravitational wave spectrum}\label{sec:reh}

If the PGW induced by inflationary models are detected, one of the important consequences would be its constraint on the reheating temperature $T_\text{th}$, which can be determined through the relation $f_\text{th}\simeq0.27\times10^{13}\,\,\text{Mpc}^{-1}\left(g_{\ast s}(T_\text{th})/106.75\right)^{1/6}\left(T_\text{th}/10^6\,\,\text{GeV}\right)$. We, therefore, aim in this section to perform the analyses on the reheating parameters for model I and model II. Since the reheating process is very sensitive to the shape of the potential as well as the coupling function, we limit our discussions only for Eqs.~(\ref{eq:potandInvGB}) and (\ref {eq:RecForm}) in this section.
The effects of the GB term during reheating epoch can be understood by how much it would change the results that are already known. By following Refs.~\cite{Dai:2014jja, Creminelli:2014fca, Munoz:2014eqa, Cai:2015soa}, we calculate the reheating parameters first. 

We start our computation by considering a mode with comoving wave number $k_\ast$ which crosses the horizon during inflation at  $a = a_\ast$. The comoving Hubble scale at the horizon crossing, $a_\ast H_\ast=k_\ast$, is related to that of the present time through the following relation:
\bea\label{eq:chzto2day}
\frac{k_\ast}{a_0H_0}&=&\frac{a_\ast}{a_\text{end}}\frac{a_\text{end}}{a_\text{th}}\frac{a_\text{th}}{a_\text{eq}}\frac{a_\text{eq}}{a_\text{0}} \frac{H_\text{eq}}{H_0}\frac{H_\ast}{H_\text{eq}}\,,
\ena
where $a_0$, $a_\ast$, $a_\text{end}$, $a_\text{th}$, and $a_\text{eq}$ denote the scale factor at the present time, time of horizon crossing, end of inflation, end of reheating, and the time of matter and radiation equality, whereas $H_0$ and $H_\text{eq}$ are the Hubble constant at the present time and the time of matter and radiation equality, respectively. We rewrite Eq.~(\ref{eq:chzto2day}) in terms of the number of $e$-folds $N$  as
\bea\label{eq:evolution}
\ln\frac{k_\ast}{a_0H_0}=-N_{_\ast}-N_\text{th}+\ln\frac{a_\text{th}}{a_0}+\ln\frac{H_\ast}{H_0}\,,
\ena
where $N_\ast\equiv \ln(a_\text{end}/a_\ast)$ is the number of $e$-folds between the time when a mode exits the horizon and the end of inflation, and $N_\text{th}\equiv \ln(a_\text{th}/a_\text{end})$ is the number of $e$-folds between the end of inflation and the end of reheating. Assuming that no entropy production took place after the completion of reheating, one can write~\cite{Dai:2014jja, Creminelli:2014fca, Munoz:2014eqa, Cai:2015soa}
\bea\label{eq:ratioatha0}
\frac{a_\text{th}}{a_0}=\frac{T_0}{T_\text{th}}\left(\frac{43}{11g_{\ast s}(T_\text{th})}\right)^{\frac{1}{3}}\,,
\ena
where $T_0$ is the current temperature of the Universe. The reheating temperature $T_\text{th}$ determines the energy density $\rho_\text{th}$ at the end of reheating:
\bea\label{eq:rhoTh}
\rho_\text{th}=\frac{\pi^2}{30}g_{\ast}(T_\text{th})T^4_{\text{th}}\,,
\ena
where $g_{\ast}(T_\text{th})$ is the number of relativistic degrees of freedom at the end of reheating. On the other hand, $\rho_\text{th}$ is related to the energy density at the end of inflation, $\rho_\text{end}$, through $N_\text{th}$  and $\omega_\text{th}$~\cite{Dai:2014jja, Creminelli:2014fca, Munoz:2014eqa}:
\bea\label{eq:rhoth}
\rho_\text{th}=\rho_\text{end} e^{-3(1+\omega_\text{th})N_\text{th}}\,.
\ena
The $\rho_\text{end}$ is a model-dependent quantity and is determined by the potential at the end of inflation $V_\text{end}$  as follows:
\bea\label{eq:rhoend}
\rho_\text{end}=\lambda_\text{end}V_\text{end}\,,
\ena
where $\lambda_\text{end}$ is an effective ratio of kinetic energy to potential energy at the end of inflation. In our case, however, it includes the effect of the Gauss-Bonnet term and is calculated as follows (see Appendix~\ref{appendix:lambdend}):
\bea\label{eq:lambend}
\lambda_\text{end}=\left.\frac{6}{6-2\epsilon-\delta_1(5-2\epsilon+\delta_2)}\right|_{\phi=\phi_\text{end}}\,.
\ena
Substituting Eqs.~(\ref{eq:rhoTh})--(\ref{eq:rhoend}) into Eq.~(\ref{eq:ratioatha0}) and then into Eq.~(\ref{eq:evolution}), we get the duration of reheating:
\bea\label{eq:Nthofth}
N_\text{th}=\frac{4}{3\omega_\text{th}-1}\left[\ln\left(\frac{k}{a_0T_0}\right) +\frac{1}{3}\ln\left(\frac{11g_{\ast s}}{43}\right)
+\frac{1}{4}\ln\left(\frac{30\lambda_\text{end}}{\pi^2g_{\ast}}\right) +\frac{1}{4}\ln\left(\frac{V_\text{end}}{H^4_\ast}\right)+N_\ast\right]\,.
\ena
With fiducial values $M_\text{pl}=\kappa^{-1}=2.435\times10^{18}$~GeV, $a_0=1$, $T_0=2.725$~K, 
$g_\ast=g_{\ast s}\simeq106.75$, and Planck's pivot scale of $k_\ast=0.05\,\,\text{Mpc}^{-1}$~\cite{Planck:2013jfk, Ade:2015lrj}, Eq.~(\ref{eq:Nthofth}) is simplified as
\bea\label{eq:NefRe}
N_\text{th}&=&\frac{4}{3\omega_\text{th}-1}\left[-60.0085
+\frac{1}{4}\ln\left(\frac{3\lambda_\text{end}}{100\pi^2}\right)
+\frac{1}{4}\ln\left(\frac{V_\text{end}}{H^4_\ast}\right)+N_\ast\right]\,,
\ena
{\color{black} where $\omega_\text{th}\neq1/3$ is assumed. If $\omega_\text{th}$ is larger (smaller) than $1/3$ in Eq.~(\ref{eq:NefRe}), the sign of the factor in front of the parentheses is positive (negative). Since $N_\text{th}\geq0$, we obtain $N_\ast\geq N_\text{extra}$ for $\omega_\text{th}>1/3$ or $N_\ast\leq N_\text{extra}$ for $\omega_\text{th}<1/3$, where 
\begin{align}
N_\text{extra} =-60.0085
+\frac{1}{4}\ln\left(\frac{3\lambda_\text{end}}{100\pi^2}\right)
+\frac{1}{4}\ln\left(\frac{V_\text{end}}{H^4_\ast}\right).
\end{align}
The expression for the reheating temperature is derived from Eqs.~(\ref{eq:rhoTh}) and (\ref{eq:rhoth}):
\bea
T_\text{th}=\left(\frac{30\lambda_\text{end}V_\text{end}}{\pi^2g_\ast}\right)^\frac{1}{4}
e^{-\frac{3}{4}(1+\omega_\text{th})N_\text{th}}\,.\label{eq:Tthofth}
\ena
The reheating temperature reaches to its maximum value if $N_\text{th}=0$ or $N_\ast=N_\text{extra}$, which implies that reheating occurs instantaneously after the end of inflation. From Eqs.~(\ref{eq:NefRe}) and (\ref{eq:Tthofth}), we see that $N_\text{th}$ and $T_\text{th}$ are linked to the inflationary quantities through $\lambda_\text{end}$, $V_\text{end}$, $N_\ast$, and $H_\ast$. These quantities need to be calculated for each model we consider in this work. 

Inflation ends when the slow-roll parameters ($\epsilon$ or $\delta_1$) become of the order of unity; $\epsilon(\phi_\text{end})\simeq1$ or $\delta_1(\phi_\text{end})\simeq1$. The substitution of Eqs.~(\ref{eq:potandInvGB}) and (\ref{eq:RecForm}) into Eqs.~(\ref{eq:srpote}) and (\ref{eq:sl_d1phi}), therefore, gives the inflaton value at the end of inflation, $\phi_\text{end}$. Once $\phi_\text{end}$ is obtained for both model I and model II, one can also calculate $V_\text{end}$ and $\lambda_{\text{end}}$ as follows:
\bea\label{eq:VendLend}
\text{model-I:} && V_\text{end}= \frac{V_0}{\kappa ^4} \left[\frac{n^2}{2}(1-\alpha)\right]^{\frac{n}{2}}\,, \quad \lambda_\text{end}=-\frac{3 n}{4 \alpha  (n+1)-2 n}\,;\\
\text{model-II:} && V_\text{end}=\frac{(\mu +x)^2}{\kappa ^4 \left(1+x^2\right)}\,,\label{eq:VandLendM2}\\
&&\lambda_\text{end}=\frac{6 \mu ^{3/2} \left(x^2+1\right)^2 \left(\sqrt{\mu }+x\right)}{6 \mu ^{3/2} \left(x^5+4 x^3+3x\right)+6 \mu ^2 \left(x^2+1\right)^2-3 \mu  \left(x^2+3\right)+\sqrt{\mu }\left(5 x^3+2x\right)+2 x^2-1}\,,\nonumber
\ena
where $x$ is a function of $\mu$ given by
\bea
x=-\frac{\sqrt{\mu }}{3}\left[1+(\mu-6)\left(\frac{2}{x_1}\right)^{\frac13}+\frac{1}{\mu}\left(\frac{2}{x_1}\right)^{-\frac13}\right]\,\label{eq:xofmu}
\ena
and
\bea
x_1=2 \mu ^3+9 \mu ^2-27 \mu+\sqrt{27(4 \mu ^5-17 \mu ^4+14 \mu ^3+27 \mu ^2)}\,.
\ena
The detailed calculations for  Eq.~(\ref{eq:VandLendM2}) are given in Appendix~\ref{appendix:phiend}. Notice from Eq.~(\ref{eq:NefRe}) that $\lambda_\text{end}$ must be positive for both models. In order for model I to yield $\lambda_\text{end}>0$, the model parameter must be in the range
\begin{align}
0\leq\alpha< \frac{n}{2+2n}.
\end{align}
In Fig.~\ref{fig:compared}, we found that the $0\leq\alpha\leq1$ range is favored by the observations at least in a $2\sigma$ confidence level. For reheating, however, the positivity of $\lambda_\text{end}$ in Eq.~(\ref{eq:NefRe}) puts another strong constraint on the model parameter.  The parameter space of $\alpha$ therefore reduces for $n>0$. Thus, we emphasize that reheating can be used as an additional constraint to the models of inflation. On the other hand, for model II, it is not simple to obtain the range of $\mu$ from Eq.~(\ref{eq:VandLendM2}) that gives $\lambda_\text{end}>0$. However, we can show the allowed range of $\mu$ numerically as is seen in Fig.~\ref{fig:lambdaii}. One can notice from Fig.~\ref{fig:lambdaii} that $\lambda_\text{end}$ diverges above a certain value of $\mu\,\, (\sim 0.3517)$.  Thus, we consider $0<\mu\leq0.3517$ in our numerical analyses in the following.
 
\begin{figure}[H]
\centering
\includegraphics[width=0.5\textwidth]{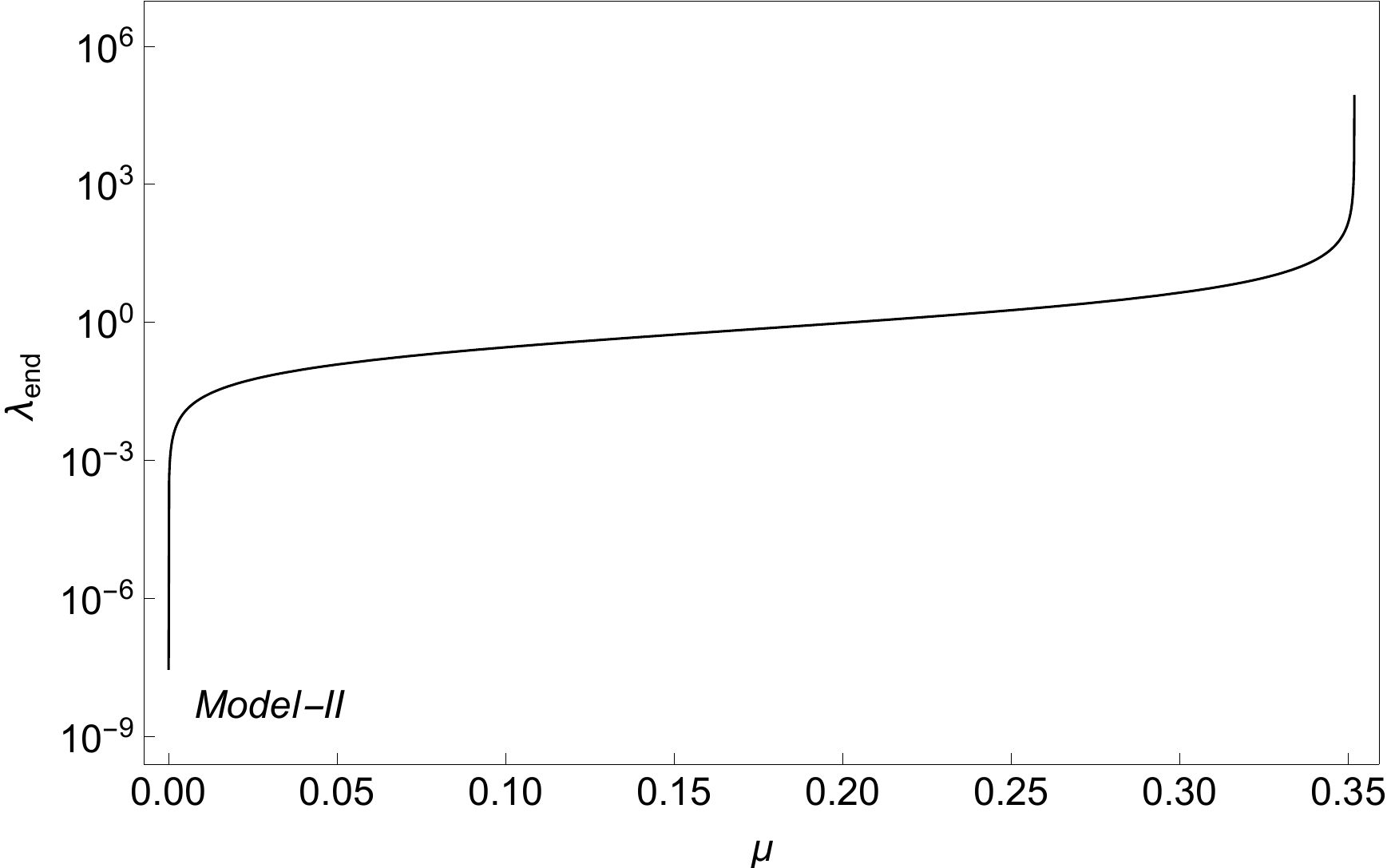}
\caption{The functional dependence of $\lambda_\text{end}$ on $\mu$ from Eq.~(\ref{eq:VandLendM2}).}\label{fig:lambdaii}
\end{figure} 

The Hubble parameter at the time of horizon crossing is obtained by substituting $\phi_\ast(N_\ast)$ into the slow-roll limit of Eq.~(\ref{beq2a}):
\bea
\text{model I:} && H_\ast^4=\left(\frac{V_0}{3\kappa^2}\right)^2\left[\frac{n}{2} (1-\alpha ) (4N_\ast+n)\right]^n\,;\label{eq:HastM1}\\
\text{model II:} && H_\ast^4=\left[\frac{(N_\ast+\mu)^2}{3\kappa^2(N_\ast^2+\mu)}\right]^2\,,\label{eq:HastM4}
\ena
where we have used from Eq.~(\ref{eq:ne}) 
\begin{align}\label{eq:NcmbM1}
\text{model I:}~~ & \kappa \phi_\ast = \sqrt{\frac{n}{2}(1-\alpha) ( 4N_\ast +n)} \,,\\
\text{model II:}~~  & \kappa \phi_\ast = \sinh^{-1} \left(\frac{1}{\sqrt{\mu}} N_{\ast} \right). \label{eq:NcmbM2}
\end{align}
In obtaining Eq.~(\ref{eq:NcmbM2}), large field inflation ($\phi_\ast\gg\phi_\text{end}$) is assumed. Here, $N_\ast$ can be expressed in terms of $n_S,\,\, n$, and $\mu$ from Eqs.~(\ref{eq:obsofN}) and~(\ref{eq:obsofNm4}).

Having obtained these quantities, we can proceed to our numerical investigation on the reheating parameters. We plot Eqs.~(\ref{eq:NefRe}) and (\ref{eq:Tthofth}) in Figs.~\ref{fig:fig4Model1}--\ref{fig:fig9Model2} for model I and model II with $\omega_\text{th}$=const. We choose four different values for $\omega_\text{th}$, namely, $\omega_\text{th}=-1/3~\text{(solid lines)}$, $0~\text{(dashed lines)}$, $1/4~\text{(dot-dashed lines)}$, and $1~\text{(dotted lines)}$. The smallest possible value for $\omega_\text{th}$ comes from the requirement that inflation has to end when $w_\text{th}=-1/3$, whereas the largest value $\omega_\text{th}=1$, the most conservative upper limit, comes from the causality. The values $\omega_\text{th}=0$ and $\omega_\text{th}=1/4$ are suggested by the literature on reheating~\cite{Munoz:2014eqa}.

We plot model I with $n=1$, $2$, and $4$ case in Fig.~\ref{fig:fig4Model1}, where the black lines indicate the absence of the GB term ($\alpha=0$) while the red ones correspond to the presence of the GB term ($\alpha\neq0$). The different curves correspond to different $\omega_\text{th}$. However, all curves intersect {\color{black} to the $N_\text{th}=0$ points, at which the instant} reheating occurs. Each dot represents different $\alpha$, and $\alpha$ increases from a black dot to a red dot. The background green shaded region corresponds to the current $1\sigma$ range $n_S=0.9655\pm0.0062$ from Planck data~\cite{Ade:2015lrj}, while the yellow band assumes the future CMB experiments with sensitivity $\pm10^{-3}$~\cite{Amendola:2012ys, Andre:2013afa}, using the same central $n_S=0.9655$ value as Planck. The horizontal blue lines at $T_\text{EW}=10^2$ GeV and $T_\text{th}=10^{6}$ GeV indicate the electroweak energy scale and the lower bound from PGW detection by DECIGO, respectively. \begin{figure}[H]
\centering
\subfigure[{\tiny $0\leq\alpha<1/4$ along the black dots.} ]
{\includegraphics[width=0.3\textwidth]{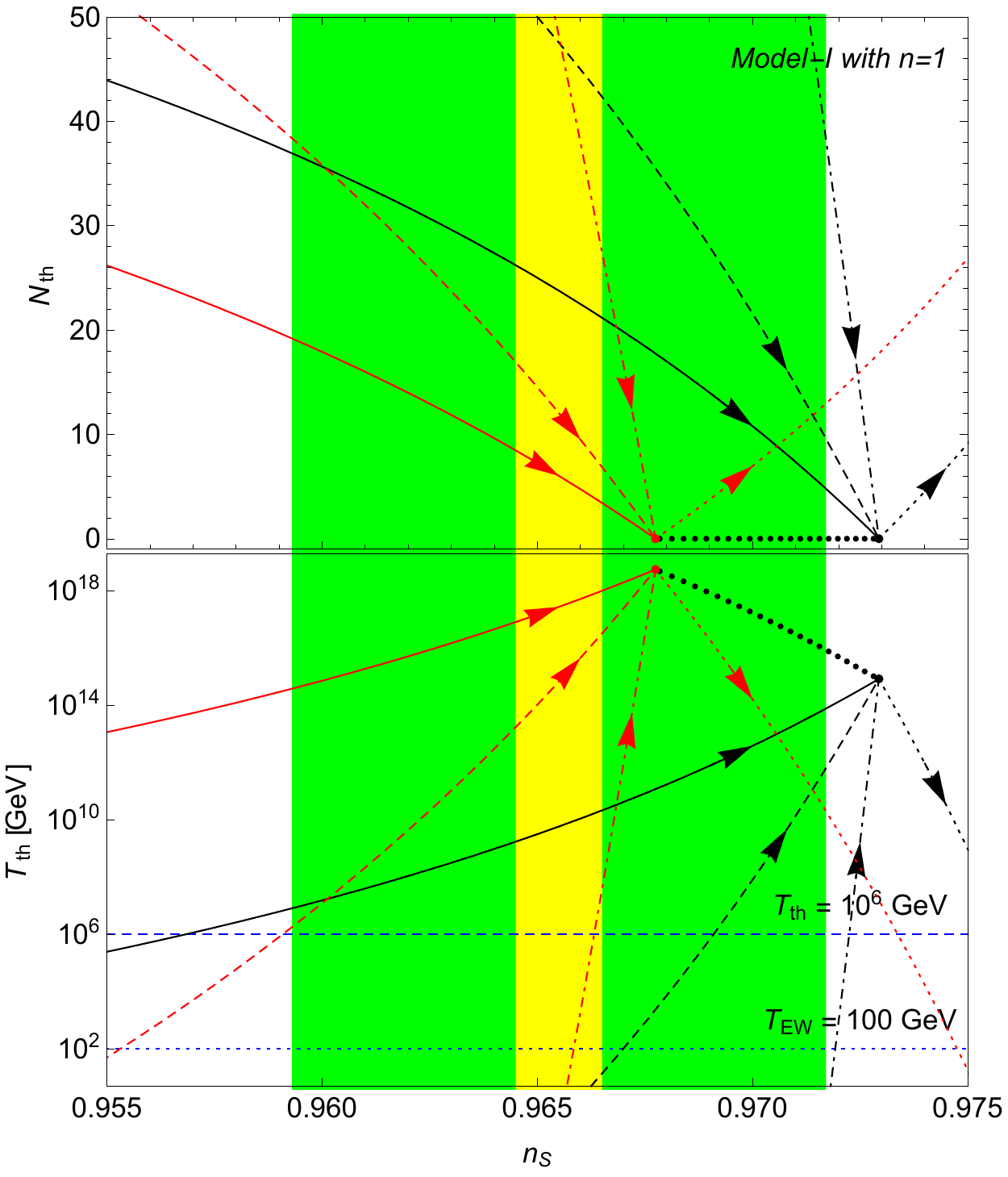}\label{fig:fig4n1}}
\subfigure[{\tiny $0\leq\alpha < 1/3$ along the black dots.}]
{\includegraphics[width=0.3\textwidth]{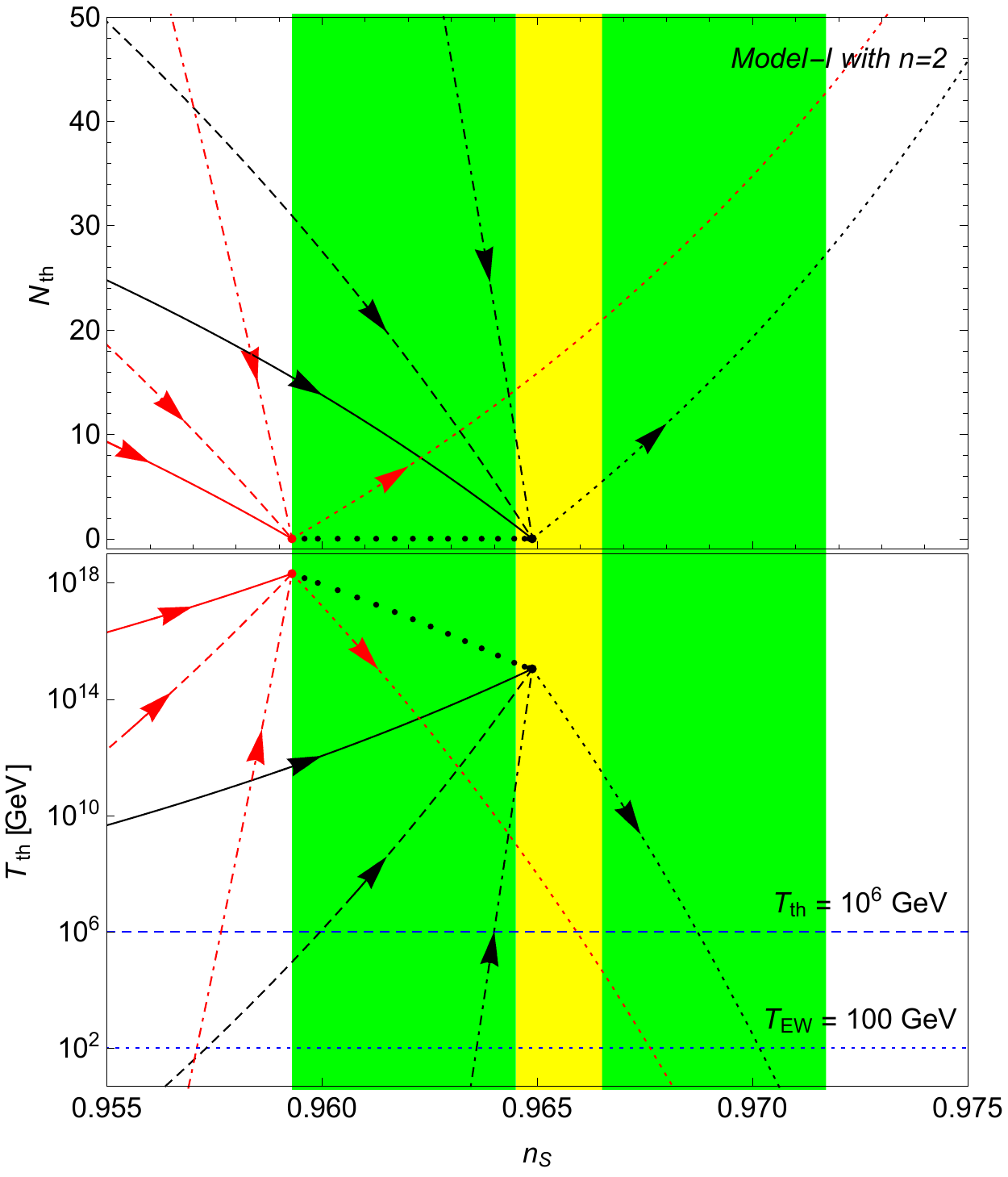}\label{fig:fig4n2}}
\subfigure[{\tiny $0\leq\alpha < 2/5$ along the black dots.}]
{\includegraphics[width=0.3\textwidth]{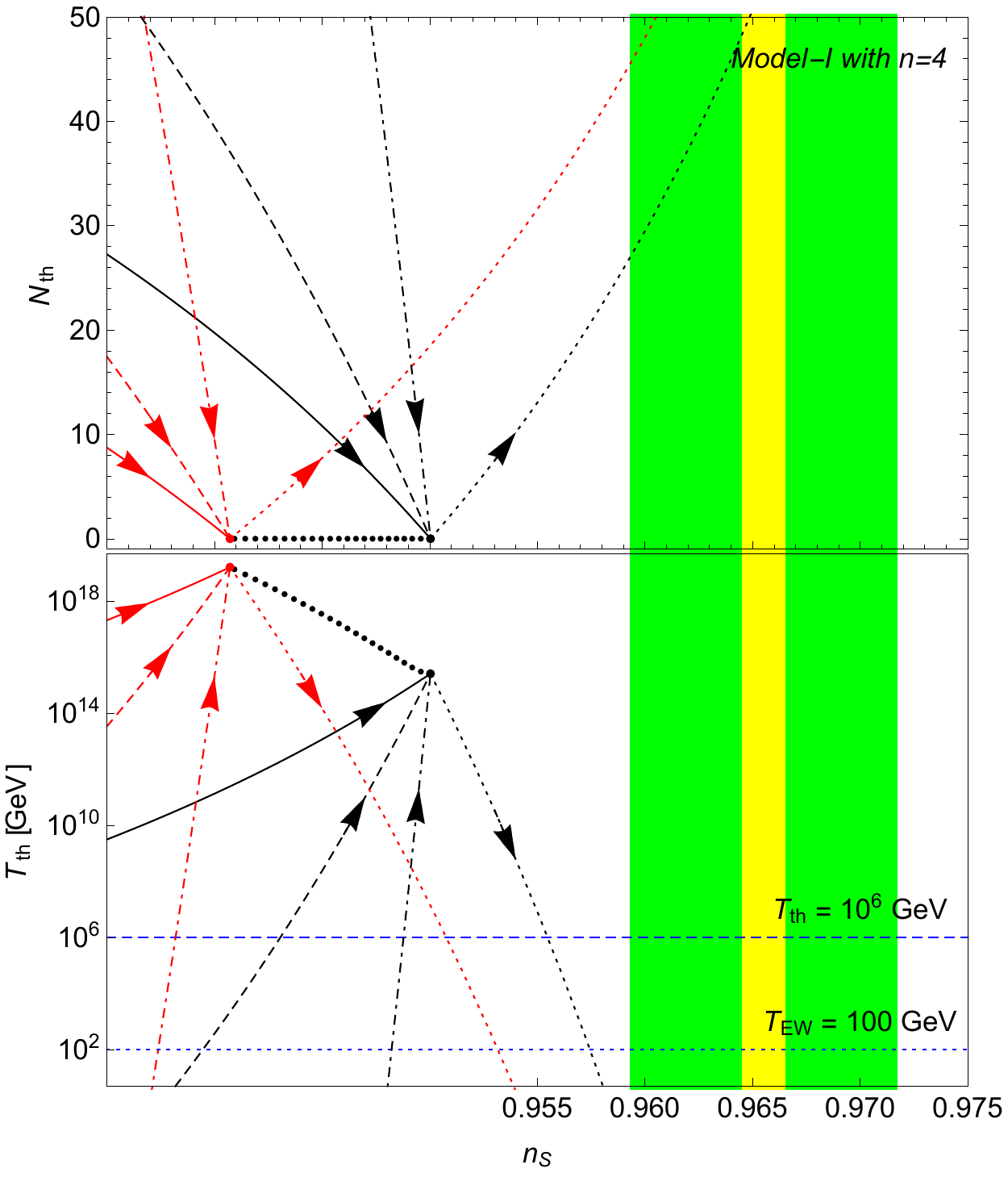}\label{fig:fig4n4}}
\caption{The $n_s$ dependence on $N_{\text{th}}$ and $T_{\text{th}}$ for model I with $V_0=0.5\times10^{-12}$. The solid black and red lines correspond to $\omega_\text{th}=-1/3$, the dashed lines to $\omega_\text{th}=0$, the dot-dashed lines to $\omega_\text{th}=1/4$, and the dotted lines to $\omega_\text{th}=1$. The black dots reaching up to the red one indicate the instantaneous reheating process with $N_\text{th}=0$ and the increasing of $\alpha$. The direction of the arrow indicates that $N_\ast$ increases along the line. The green shaded region corresponds to the current $1\sigma$ range $n_S=0.9655\pm0.0062$ from Planck data~\cite{Ade:2015lrj}, while the yellow band assumes the future CMB experiments with sensitivity $\pm10^{-3}$~\cite{Amendola:2012ys, Andre:2013afa}, using the same central $n_S=0.9655$ value as Planck. The horizontal blue lines at $T_\text{EW}=10^2$ GeV (dotted) and $T_\text{th}=10^{6}$ GeV (dashed) indicate the electroweak (EW) scale and the lower bound from PGW detection by DECIGO, respectively.}\label{fig:fig4Model1}
\end{figure}
The arrows in Fig.~\ref{fig:fig4Model1} indicate that $N_\ast$ increases along the line. The direction of the arrow is determined by the sign of the factor in front of the square bracket in Eq.~(\ref{eq:NefRe}). They move toward the point of $N_\text{th}=0$ for $\omega_\text{th}<1/3$ and move away for $\omega_\text{th} > 1/3$. Since $N_\text{th}$ indicates the duration of reheating, it must be positive or zero. To yield $N_\text{th}\geq0$, $N_\ast$ must be smaller than $N_\text{extra}$ in Eq.~(\ref{eq:NefRe}) for $\omega_\text{th}<1/3$ {\color{black} and be larger for $\omega_\text{th} > 1/3$.} Thus, the direction of the arrow indicates the increasing of $N_\ast$ during inflation. One can notice that $N_\ast$ increases as $n_s$ increases for a given $n$ and $\alpha$ from Eq.~(\ref{eq:obsofN}).
\begin{figure}[H]
\centering
\subfigure[]
{\includegraphics[width=0.4\textwidth]{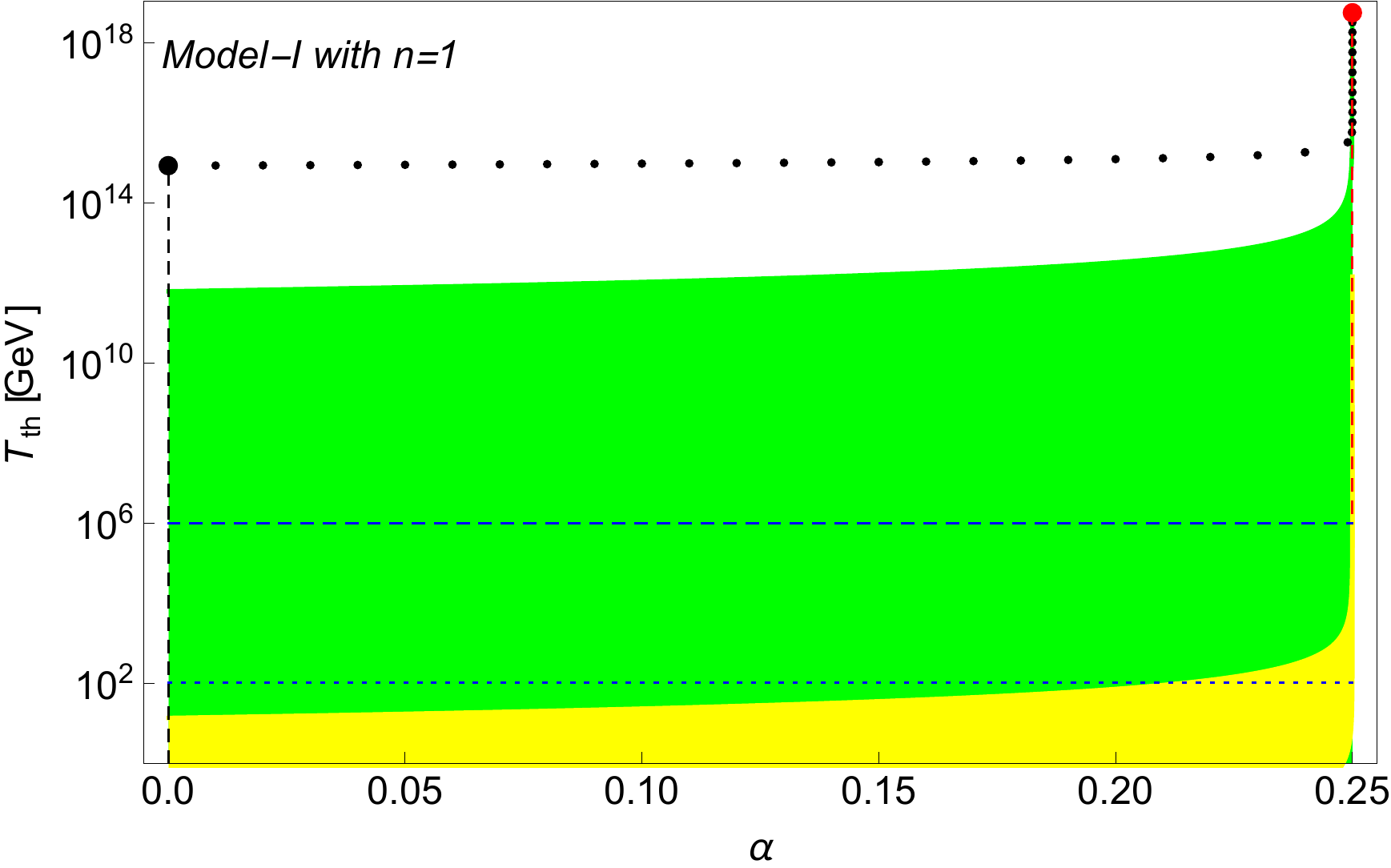}\label{fig:fig6n1}}
\subfigure[]
{\includegraphics[width=0.4\textwidth]{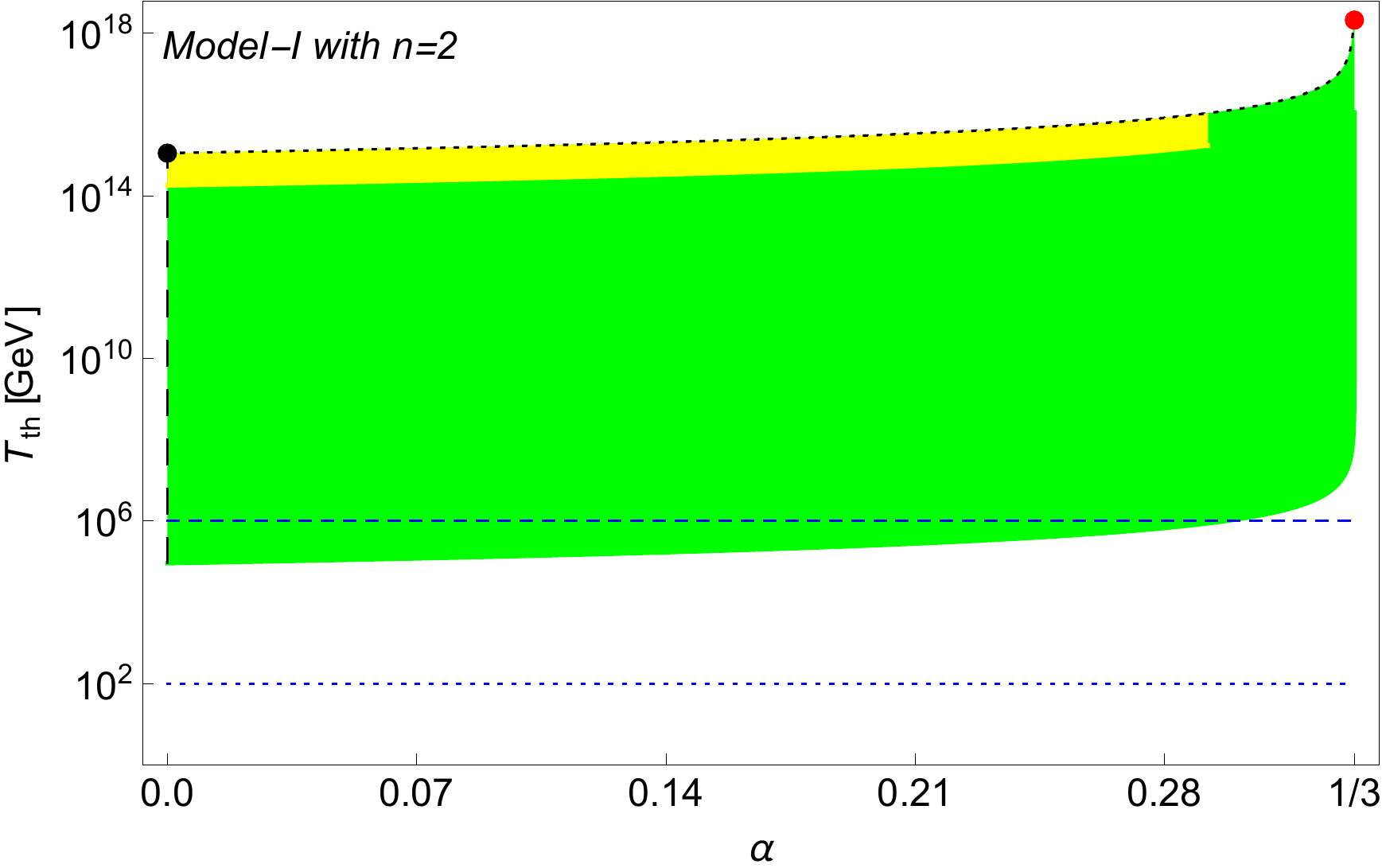}\label{fig:fig6n2}}
\caption{The reheating temperature $T_\text{th}$ as a function of $\alpha$ when $\omega_\text{th}=0$. The vertical black and red dashed lines correspond to $\omega_\text{th}=0$ dashed lines in Fig.~\ref{fig:fig4Model1}. The black dots reaching up to the red one indicate the instantaneous reheating with $N_\text{th}=0$. The background shared regions, as well as the horizontal lines, are as for Fig.~\ref{fig:fig4Model1}.}\label{fig:fig6Model1}
\end{figure}
The reheating temperature appears to be significantly increasing as $\alpha$ increases. In Fig.~\ref{fig:fig6Model1}, therefore, we plot the $\alpha$ dependence of $T_\text{th}$ from Eq.~(\ref{eq:Tthofth}) {\color{black} with $\omega_\text{th} =0$ for $n=1$ and $2$}. The black vertical dashed lines at $\alpha =0$ and the red vertical dashed lines at  $1/4$ and $1/3$ correspond to the same $\omega_\text{th}=0$ dashed lines in Fig.~\ref{fig:fig4n1} and \ref{fig:fig4n2}, respectively. There is no vertical red line in Fig.~\ref{fig:fig6n2}, because the red point in Fig.~\ref{fig:fig4n2} locates at the boundary of the $1\sigma$ region. When $\alpha=0$, the reheating temperature peaks at $T_\text{th}\sim10^{15}$ GeV in each of the three cases; see the bigger black intersecting points in Figs.~\ref{fig:fig4Model1} and~\ref{fig:fig6Model1}. When $\alpha\neq0$, the maximum $T_\text{th}$ is denoted by the red points, but the exact values depend on $\alpha$ for each $n$. 

Similar results are also obtained for model II. Although the wide range of $\mu$ is acceptable, the reliable range must be given from the condition $\lambda_\text{end}>0$. From Fig.~\ref{fig:lambdaii}, we obtained 
$0 < \mu \leq 0.3517$. With this reliable range of $\mu$, we plot Fig.~\ref{fig:fig7Model2}. Together with Fig.~\ref{fig:fig9Model2} where $\omega_\text{th} =0$, it shows that the reheating temperature is increasing as $\mu$ increases.  The result is valid for other constant values of $\omega_\text{th}$.
\begin{figure}[H]
\centering
\includegraphics[width=0.491\textwidth]{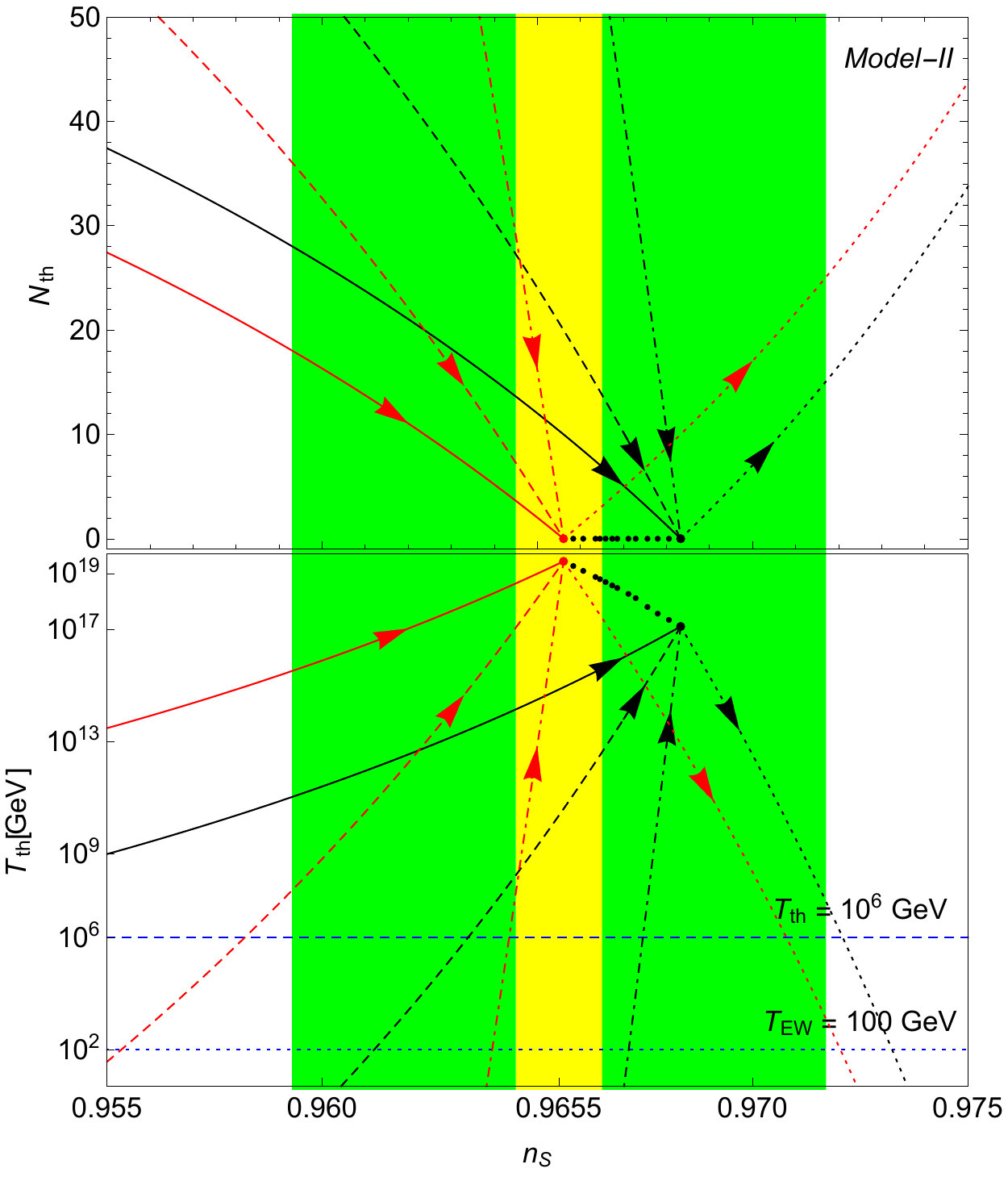}\label{fig:fig7sub}
\caption{The $n_s$ dependence on $N_{\text{th}}$ and $T_{\text{th}}$ for model II with $V_0=0.5\times10^{-12}$. The solid black and red lines correspond to $\omega_\text{th}=-1/3$, the dashed lines to $\omega_\text{th}=0$, the dot-dashed lines to $\omega_\text{th}=1/4$, and the dotted lines to $\omega_\text{th}=1$. The black dots reaching up to the red one indicate the instantaneous reheating process with $N_\text{th}=0$ and the increasing of $\mu$ between $10^{-4}\leq\mu\leq 0.3517$. The direction of the arrow indicates that $N_\ast$ increases along the line. The shaded regions, as well as the horizontal lines, are same as for Fig.~\ref{fig:fig4Model1}.}\label{fig:fig7Model2}
\end{figure}
\begin{figure}[H]
\centering
{\includegraphics[width=0.491\textwidth]{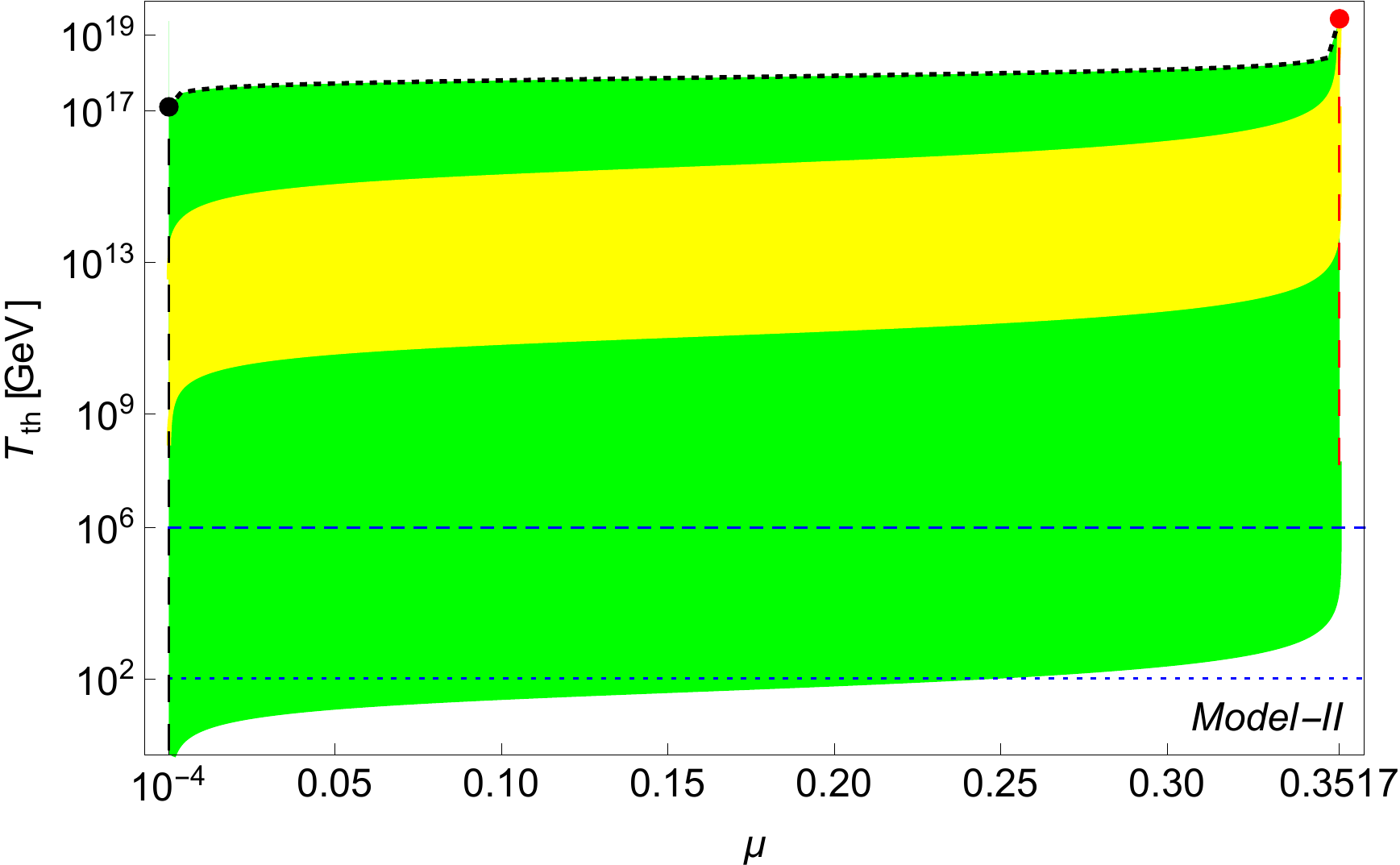}}
\caption{The black and red dots, respectively, at $(\mu, T_\text{th}) = (10^{-4}, 1.27\times10^{17}\,\,\text{GeV})$ and $(0.3517, 2.69\times10^{19}\,\,\text{GeV})$ correspond to the maximum reheating temperatures for instantaneous reheating ($N_\text{th}=0$). The vertical black and red dashed lines represent $\omega_\text{th}=0$ lines in Fig.~\ref{fig:fig7Model2}. The shaded regions, as well as the horizontal lines, are same as for Fig.~\ref{fig:fig4Model1}.}\label{fig:fig9Model2}
\end{figure}
Our results therefore imply that the presence of the GB term during inflation significantly enhances the thermalization temperature at the end of reheating. Once the reheating temperature is determined by the detection of PGW, other parameters including $\omega_\text{th}$ and $N_\text{th}$ can also be determined with a help of Figs.~\ref{fig:fig4Model1} and~\ref{fig:fig7Model2} in light of the current or future observational data.

\section{Conclusion}\label{sec:conc}
After a brief review on the basics, we discuss two types of inflation model with the GB term~\cite{Koh:2014bka, Jiang:2013gza, Koh:2016abf, Guo:2010jr}: the models that predict a red-tilted ($n_T<0$) primordial tensor power spectrum (model I) given in Eq.~(\ref{eq:potandInvGB}) and those that predict a blue-tilted ($n_T>0$) spectrum (model II) given in Eq.~(\ref{eq:RecForm}). For model I and model II, we estimated the energy spectrum of the PGW and the reheating parameters after GB inflation.

The expression for the energy spectrum of the PGW background is calculated in Eq.~(\ref{ESofGWofk}), which is the main analytic result of Sec.~\ref{sec:GWES}, and is plotted as a function of the frequency in Fig.~\ref{fig:GWconst} for model I and model II. The amplitude of the PGWs for model I is suppressed for increasing values of $\alpha$, because the inflationary tensor power spectrum given in Eq.(\ref{eq:tpowspec}) has a red tilt for the positive $\alpha$.  On the other hand, the amplitude for model II is significantly enhanced as the model parameter $\mu$ increases, in which the inflationary tensor power spectrum is predicted to have a blue tilt. Moreover, if the running of the tensor spectral index is considered in the estimation, the spectrum of model I is more suppressed, whereas that of model II is more enhanced as the frequency increases; see Figs.~\ref{fig:fig2c} and~\ref{fig:fig2d}. This implies $\alpha_T < 0$ for model I, but $\alpha_T >0$ for model II. This indicates the running of the tensor spectral index found to be an important quantity for the detectability of the PGW background if the inflationary tensor power spectrum have a blue tilt, for example, in the 0.1--1 Hz range by DECIGO~\cite{Kudoh:2005as}.

It is worth noting from Eqs.~(\ref{eq:obsofN}) and (\ref{eq:obsofNm4}) that the tensor-to-scalar ratio is inversely proportional to the number of $e$-folds for model I, whereas it is inversely proportional to the square of the number of $e$-folds for model II. Therefore, the expected GW signal in the region of interferometers is not dramatically different for the two models we consider in this study. However, we showed that the small changes in model parameters $\alpha$ and $\mu$ significantly modify the GW energy spectrum. The values of $\alpha$ and $\mu$ within the range $0\leq\alpha<1$ for model I and $0<\mu\lesssim 50$ for model II seem to provide detectable GW signals by the DECIGO experiment. For other models that belong to either of the two categories but have not been discussed in this work, there could be a possible detection by other ground- or space-based experiments such as LIGO/VIRGO and LISA or pulsar timing array experiments~\cite{AmaroSeoane:2012km, Ferdman:2010xq, Kudoh:2005as}.

An important consequence of Sec.~\ref{sec:GWES} is that the reheating temperature can be determined by the frequency at which the PGW background would be detected. If signals of the PGW induced by model I and model II are detected, the bound on the $T_\text{th}$ can be read off from Fig.~\ref{fig:GWconst}. Since the signals of the PGW produced by both model I and model II are observable by DECIGO around $f\simeq0.1$--$10$ Hz, the reheat temperature would be $T_\text{th} \gtrsim 10^{6}$ GeV. Once the thermalization temperature at the end of reheating is determined, the other reheating parameters (the duration $N_\text{th}$ and the equation of state $\omega_\text{th}$) could also be determined.

In Sec.~\ref{sec:reh}, we derived the reheating parameters after Gauss-Bonnet inflation in Eqs.~(\ref{eq:NefRe}) and~(\ref{eq:Tthofth}) and showed that the reheating parameters are highly sensitive to the presence of the GB term for both models. By assuming the constant equation-of-state $\omega_\text{th}$ parameter during reheating, and no entropy production took place after reheating, we numerically estimated the reheating parameters using our analytic results in Figs.~\ref{fig:fig4Model1} and~\ref{fig:fig7Model2}. As the figures show, the thermalization temperature at the end of reheating is significantly enhanced for both models due to the presence of the GB term. Moreover, the duration of reheating can be read off from the upper panels of Figs.~\ref{fig:fig4Model1} and~\ref{fig:fig7Model2} if the temperature $T_\text{th}$ is determined. $T_\text{th}$ is assumed to be determined by the detection of the PGW in this work. Both model I and model II support the equation-of-state parameter between $-1/3\leq\omega_\text{th}\leq1$. In the preferred range of $\omega_\text{th}$, the physically meaningful range of temperature from $\sim 100$ GeV to around $\sim 10^{16}$ GeV can be safely accommodated. An important consequence of this section that we emphasize is that reheating can be used as an additional constraint to the models of inflation and can reduce the parameter space.

We have reached the conclusion that the GB term seems to be important not only during inflation but also during reheating whether the process is instantaneous or lasts for a certain number of $e$-folds until it completes. Once $T_\text{th}$ is determined, the other reheating parameters can be estimated in our models. Thus, it would be interesting to investigate the physics of reheating in light of forthcoming GW experiments.

\begin{acknowledgments}
We thank Ryusuke Jinno, Kohei Kamada, Sachiko Kuryanagi, and Massimo Giovannini for their helpful comments. The work of G. T. was supported by IBS under the project code IBS-R018-D1. B. -H. L. was supported by the Basic Science Research Program through the National Research Foundation of Korea funded by the Ministry of Education (No. NRF-2018R1D1A1B07048657). S. K. was supported by the Basic Science Research Program through the National Research Foundation of Korea funded by the Ministry of Education (No. NRF-2016R1D1A1B04932574). 
\end{acknowledgments}

\appendix
\section{THE CALCULATION ON $\lambda_\text{end}$}\label{appendix:lambdend}
In this Appendix, we derive the expression given in Eq.~(\ref{eq:lambend}). We start our calculation with the energy density of inflaton, which can be read off from Eq.~(\ref{beq2a}), at the end of inflation,
\bea\label{eq:rhoGB}
\rho_\text{end}=\left. \left(\frac{1}{2}\dot{\phi}^2+V+12\dot{\xi}H^3\right)\right|_{\phi=\phi_\text{end}}=\lambda_\text{end} V_\text{end}\,,
\ena
where $\lambda_\text{end}$ is defined as follows:
\bea\label{eq:lambdend}
\lambda_\text{end}\equiv\left.\left(\frac{1}{2}\frac{\dot{\phi}^2}{V}+1+\frac{12\dot{\xi}H^3}{V}\right)\right|_{\phi=\phi_\text{end}}\,.
\ena
On the other hand, the potential energy can be obtained from Eq.~(\ref{beq2a}) as
\bea\label{eq:potofH}
V=\frac{3}{\kappa^2}H^2-\frac{1}{2}\dot{\phi}^2-12\dot{\xi}H^3\,.
\ena
Substituting Eq.~(\ref{eq:potofH}) into Eq.~(\ref{eq:lambdend}), we get the $\lambda_{\text{end}}$ which is given by
\bea\label{eq:lambdaend}
\lambda_{\text{end}}&\equiv&
\left. \left[1+\left(\frac{6H^2}{\kappa^2\dot{\phi}^2}-1-\frac{24\kappa^2\dot{\xi}H^3}{\kappa^2\dot{\phi}^2}\right)^{-1} +\left(\frac{1}{4\kappa^2\dot{\xi}H} - \frac{\kappa^2\dot{\phi}^2}{24\kappa^2\dot{\xi}H^3}-1\right)^{-1}\right]\right|_{\phi=\phi_\text{end}}\,.
\ena
Here, we use Eq.~(\ref{eq:sl_param}) together with the following equation from Eq.~(39) of Ref.~\cite{Koh:2014bka}:
\bea
\frac{\kappa^2\dot{\phi}^2}{H^2}=2\epsilon-\delta_1(1+2\epsilon-\delta_2)\,,
\ena
to express Eq.~(\ref{eq:lambdaend}) in terms of the slow-roll parameters:
\bea
\lambda_{\text{end}}
=\left. \frac{6}{6-2\epsilon-\delta_1(5-2\epsilon+\delta_2)}\label{eq:lambda}\right|_{\phi=\phi_\text{end}}\,.
\ena
When $\delta_1=0$ in Eq.~(\ref{eq:lambda}), the Gauss-Bonnet term becomes absent, which shows the consistency between our result and that of Refs.~\cite{Dai:2014jja, Creminelli:2014fca, Munoz:2014eqa}.

\section{THE CALCULATION ON $\phi_\text{end}$ FOR MODEL II}\label{appendix:phiend}

Inflation ends when either one of the slow-roll parameters in Eqs.~(\ref{eq:srpote})--(\ref{eq:sl_d2phi}) becomes of the order of unity. For example, $\epsilon(\phi_\text{end})\equiv1$ at the end of inflation.
For model II, that has the potential and the coupling functions of the form given in Eq.~(\ref{eq:RecForm}), we obtain
\bea\label{eq:epsofphiend}
\epsilon(\phi)&=&\frac{\text{sech}^2(\kappa \phi)\left[1-\sqrt{\mu }\sinh(\kappa  \phi )\right]}{\mu+\sqrt{\mu } \sinh (\kappa  \phi ) }\,.
\ena
By defining a new variable $x\equiv\sinh(\kappa \phi)$ in Eq.~(\ref{eq:epsofphiend}), we can write $\epsilon(\phi_\text{end})=1$ as follows:
\bea
\sqrt{\mu } x^3+\mu  x^2+2 \sqrt{\mu } x+\mu-1=0\,.
\ena
This equation has a real solution of the form
\bea
x=-\frac{\sqrt{\mu }}{3}\left[1+(\mu-6)\left(\frac{2}{x_1}\right)^{\frac13}+\frac{1}{\mu}\left(\frac{2}{x_1}\right)^{-\frac13}\right]\,,\label{eq:xofmu}
\ena
where
\bea
x_1=2 \mu ^3+9 \mu ^2-27 \mu+\sqrt{27(4 \mu ^5-17 \mu ^4+14 \mu ^3+27 \mu ^2)}\,.
\ena
It is worth noting that $x_1$ is positive for $\mu>0$. However, $x$ is positive for $0<\mu<1$ and is negative for $\mu>1$. When $\mu=1$, we have $x=0$.

Solving $\sinh(\kappa\phi_\text{end})=x$ for $\kappa\phi_\text{end}$, we find the inflaton field value at the end of inflation as
\bea
\kappa\phi_\text{end}=-\text{arcsinh}\left[\frac{\sqrt{\mu }}{3}\left(1+(\mu-6)\left(\frac{2}{x_1}\right)^{\frac13}+\frac{1}{\mu}\left(\frac{2}{x_1}\right)^{-\frac13}\right)\right]+2\pi i c_1\,,
\ena
where $c_1$ is an arbitrary constant.
The potential energy at the end of inflation therefore becomes
\bea
V_\text{end}=\frac{1}{\kappa^4}\frac{(\mu+x)^2}{1+x^2}\,,
\ena
and $\lambda_\text{end}$ gets
\bea\label{eq:lambdaendmu}
\lambda_\text{end}=\frac{6 \mu ^{3/2} \left(x^2+1\right)^2 \left(\sqrt{\mu }+x\right)}{6 \mu ^2 \left(x^2+1\right)^2-3 \mu  \left(x^2+3\right)+\sqrt{\mu } x \left(5 x^2+2\right)+2 x^2+6 \mu ^{3/2} x \left(x^4+4 x^2+3\right)-1}\,.\nonumber\\
\ena
Since $x$ is given in Eq.~(\ref{eq:xofmu}) as a function of $\mu$, both $V_\text{end}$ and $\lambda_\text{end}$ are functions only of $\mu$. As we mentioned earlier, $\lambda_\text{end}$ must be positive in order to yield $N_\text{th}\geq0$. Thus, in Fig.~\ref{fig:lambdaii}, we plot the positive range of $\lambda_\text{end}$ as a function of $\mu$. We see that $\lambda_\text{end}$ diverges around $\mu\simeq0.3517$.

\bibliography{draft}

\providecommand{\href}[2]{#2}\begingroup\raggedright\begin{thebibliography}{10}
\bibitem{Guth:1980zm}
  A.~H.~Guth, The inflationary universe: A possible solution to the horizon and flatness problems,''
  Phys.\ Rev.\ D {\bf 23}, 347 (1981).

\bibitem{Albrecht:1982wi}
  A.~Albrecht and P.~J.~Steinhardt,
  Cosmology for Grand Unified Theories with Radiatively Induced Symmetry Breaking,
  Phys.\ Rev.\ Lett.\  {\bf 48}, 1220 (1982).

\bibitem{Linde:1981mu}
  A.~D.~Linde,
  A new inflationary universe scenario: A possible solution of the horizon, flatness, homogeneity, isotropy and primordial monopole problems,''
  Phys.\ Lett.\ {\bf 108 B}, 389 (1982).


\bibitem{Hinshaw:2012aka}
  G.~Hinshaw {\it et al.} (WMAP Collaboration),
  Nine-year Wilkinson microwave anisotropy probe observations: Cosmological parameter results,''
  Astrophys.\ J.\ Suppl.\  {\bf 208}, 19 (2013);

\bibitem{Komatsu:2010fb}
  E.~Komatsu {\it et al.} (WMAP Collaboration),
  Seven-year Wilkinson microwave anisotropy probe observations: Cosmological interpretation,
  Astrophys.\ J.\ Suppl.\  {\bf 192}, 18 (2011)\,.

\bibitem{Komatsu:2008hk}
  E.~Komatsu {\it et al.} (WMAP Collaboration),
  Five-year Wilkinson microwave anisotropy probe observations: Cosmological interpretation,
  Astrophys.\ J.\ Suppl.\  {\bf 180}, 330 (2009)\,.

\bibitem{Planck:2013jfk}
  P.~A.~R.~Ade {\it et al.} (Planck Collaboration),
  Planck 2013 results. XXII. Constraints on inflation,
  Astron.\ Astrophys.\  {\bf 571}, A22 (2014)\,.

\bibitem{Ade:2015lrj}
  P.~A.~R.~Ade {\it et al.} (Planck Collaboration),
  Planck 2015 results. XX. Constraints on inflation,
  Astron.\ Astrophys.\  {\bf 594}, A20 (2016)\,.

\bibitem{Starobinsky:1979ty}
  A.~A.~Starobinsky,
  Spectrum of relict gravitational radiation and the early state of the universe,
  Pis'ma Zh.\ Eksp.\ Teor.\ Fiz.\  {\bf 30}, 719 (1979)
  [JETP Lett.\  {\bf 30}, 682 (1979)]\,.

\bibitem{Allen:1987bk}
  B.~Allen,
  The stochastic gravity wave background in inflationary universe models,
  Phys.\ Rev.\ D {\bf 37}, 2078 (1988).

 \bibitem{Sahni:1990tx}
  V.~Sahni,
  The energy density of relic gravity waves from inflation,
  Phys.\ Rev.\ D {\bf 42}, 453 (1990).

\bibitem{Kamionkowski:1996ks}
  M.~Kamionkowski, A.~Kosowsky, and A.~Stebbins,
  Statistics of cosmic microwave background polarization,
  Phys.\ Rev.\ D {\bf 55}, 7368 (1997)\,.

\bibitem{AmaroSeoane:2012km}
  P.~Amaro-Seoane {\it et al.},
  eLISA/NGO: Astrophysics and cosmology in the gravitational-wave millihertz regime,
  arXiv:1201.3621.

\bibitem{Ferdman:2010xq}
  R.~D.~Ferdman {\it et al.},
  The European pulsar timing array: Current efforts and a LEAP toward the future,
  Classical \ Quantum\ Gravity\  {\bf 27}, 084014 (2010)\,.

\bibitem{Kudoh:2005as}
  H.~Kudoh, A.~Taruya, T.~Hiramatsu, and Y.~Himemoto,
  Detecting a gravitational-wave background with next-generation space interferometers,
  Phys.\ Rev.\ D {\bf 73}, 064006 (2006)\,;

  S.~Kawamura {\it et al.},
  The Japanese space gravitational wave antenna: DECIGO,
  Classical\ Quantum\ Gravity\  {\bf 28}, 094011 (2011).

\bibitem{Turner:1993vb}
  M.~S.~Turner, M.~J.~White, and J.~E.~Lidsey,
  Tensor perturbations in inflationary models as a probe of cosmology,
  Phys.\ Rev.\ D {\bf 48}, 4613 (1993)\,.

\bibitem{Zhao:2006mm}
  W.~Zhao and Y.~Zhang,
  Relic gravitational waves and their detection,
  Phys.\ Rev.\ D {\bf 74}, 043503 (2006)\,.

\bibitem{Watanabe:2006qe}
  Y.~Watanabe and E.~Komatsu,
  Improved calculation of the primordial gravitational wave spectrum in the standard model,
  Phys.\ Rev.\ D {\bf 73}, 123515 (2006)\,.

\bibitem{Nakayama:2008wy}
  K.~Nakayama, S.~Saito, Y.~Suwa, and J.~Yokoyama,
  Probing reheating temperature of the universe with gravitational wave background,
  J. Cosmol. Astropart. Phys. {\bf 06} (2008) 020\,.

\bibitem{Kuroyanagi:2008ye}
  S.~Kuroyanagi, T.~Chiba, and N.~Sugiyama,
  Precision calculations of the gravitational wave background spectrum from inflation,
  Phys.\ Rev.\ D {\bf 79}, 103501 (2009)\,.

\bibitem{Giovannini:2008tm}
  M.~Giovannini,
  Thermal history of the plasma and high-frequency gravitons,
  Classical\ Quantum\ Gravity\  {\bf 26}, 045004 (2009)\,.

\bibitem{Nakayama:2009ce}
  K.~Nakayama and J.~Yokoyama,
  Gravitational wave background and non-Gaussianity as a probe of the curvaton scenario,''
  J. Cosmol. Astropart. Phys. {\bf 01} (2010) 010\,.

\bibitem{Kuroyanagi:2014nba}
  S.~Kuroyanagi, T.~Takahashi, and S.~Yokoyama,
  Blue-tilted tensor spectrum and thermal history of the universe,
  J. Cosmol. Astropart. Phys. {\bf 02} (2015) 003\,.

\bibitem{Tumurtushaa:2016ars}
  G.~Tumurtushaa, S.~Koh, and B.~H.~Lee,
  Suppression of the primordial gravitational waves,
  Int.\ J.\ Mod.\ Phys.\ Conf.\ Ser.\  {\bf 43}, 1660204 (2016).

\bibitem{Koh:2015brl}
  S.~Koh, B.~H.~Lee, and G.~Tumurtushaa,
  Primordial gravitational waves from the space-condensate inflation model,
  Phys.\ Rev.\ D {\bf 93}, 083518 (2016)\,.

\bibitem{Kuroyanagi:2011fy}
  S.~Kuroyanagi, K.~Nakayama, and S.~Saito,
  Prospects for determination of thermal history after inflation with future gravitational wave detectors,
  Phys.\ Rev.\ D {\bf 84}, 123513 (2011)\,.

\bibitem{Buchmuller:2013lra}
  W.~Buchmller, V.~Domcke, K.~Kamada, and K.~Schmitz,
  The gravitational wave spectrum from cosmological $B-L$ breaking,
  J. Cosmol. Astropart. Phys. {\bf 10} (2013) 003\,.

\bibitem{Nojiri:2005vv} 
  S.~Nojiri, S.~D.~Odintsov, and M.~Sasaki,
  Gauss-Bonnet dark energy,
  Phys.\ Rev.\ D {\bf 71}, 123509 (2005)\,.

\bibitem{Satoh:2010ep}
  M.~Satoh,
  Slow-roll inflation with the Gauss-Bonnet and Chern-Simons corrections,
  J. Cosmol. Astropart. Phys. {\bf 11} (2010) 024.
  
\bibitem{Satoh:2008ck} 
  M.~Satoh and J.~Soda,
  Higher curvature corrections to primordial fluctuations in slow-roll inflation,
  J. Cosmol. Astropart. Phys. {\bf 09} (2008) 019.

\bibitem{Guo:2010jr}
  Z.~K.~Guo and D.~J.~Schwarz,
  Slow-roll inflation with a Gauss-Bonnet correction,
  Phys.\ Rev.\ D {\bf 81}, 123520 (2010)\,.

\bibitem{Chakraborty:2018scm} 
  S.~Chakraborty, T.~Paul, and S.~SenGupta,
  Inflation driven by Einstein-Gauss-Bonnet gravity,
  arXiv:1804.03004.

\bibitem{Koh:2014bka}
  S.~Koh, B.~H.~Lee, W.~Lee, and G.~Tumurtushaa,
  Observational constraints on slow-roll inflation coupled to a Gauss-Bonnet term,
  Phys.\ Rev.\ D {\bf 90}, 063527 (2014).

\bibitem{Jiang:2013gza}
  P.~X.~Jiang, J.~W.~Hu, and Z.~K.~Guo,
  Inflation coupled to a Gauss-Bonnet term,
  Phys.\ Rev.\ D {\bf 88}, 123508 (2013)\,.

\bibitem{Koh:2016abf}
  S.~Koh, B.~H.~Lee, and G.~Tumurtushaa,
  Reconstruction of the scalar field potential in inflationary models with a Gauss-Bonnet term,
  Phys.\ Rev.\ D {\bf 95}, 123509 (2017)\,.

\bibitem{Satoh:2007gn}
  M.~Satoh, S.~Kanno, and J.~Soda,
  Circular polarization of primordial gravitational waves in string-inspired inflationary cosmology,
  Phys.\ Rev.\ D {\bf 77}, 023526 (2008)\,.

\bibitem{vandeBruck:2016xvt}
  C.~van de Bruck, K.~Dimopoulos, and C.~Longden,
  Reheating in Gauss-Bonnet-coupled inflation,
  Phys.\ Rev.\ D {\bf 94}, 023506 (2016)\,.

\bibitem{Bhattacharjee:2016ohe}
  S.~Bhattacharjee, D.~Maity, and R.~Mukherjee,
  Constraining scalar-Gauss-Bonnet inflation by reheating, unitarity and PLANCK,
  Phys.\ Rev.\ D {\bf 95}, 023514 (2017)\,.


\bibitem{Nozari:2017rta}
  K.~Nozari and N.~Rashidi,
  Perturbation, non-Gaussianity, and reheating in a Gauss-Bonnet $\alpha$-attractor model,
  Phys.\ Rev.\ D {\bf 95}, 123518 (2017)\,.

\bibitem{Dai:2014jja}
  L.~Dai, M.~Kamionkowski, and J.~Wang,
  Reheating Constraints to Inflationary Models,
  Phys.\ Rev.\ Lett.\  {\bf 113}, 041302 (2014)\,.

\bibitem{Creminelli:2014fca}
  P.~Creminelli, D.~L\'{o}pez Nacir, M.~Simonovi\'{c}, G.~Trevisan, and M.~Zaldarriaga,
  $\phi^2$ inflation at its endpoint,
  Phys.\ Rev.\ D {\bf 90}, 083513 (2014)\,.

\bibitem{Munoz:2014eqa}
  J.~B.~Munoz and M.~Kamionkowski,
  Equation-of-state parameter for reheating,
  Phys.\ Rev.\ D {\bf 91}, 043521 (2015)\,.

\bibitem{Cai:2015soa}
  R.~G.~Cai, Z.~K.~Guo, and S.~J.~Wang,
  Reheating phase diagram for single-field slow-roll inflationary models,
  Phys.\ Rev.\ D {\bf 92}, 063506 (2015)\,.

\bibitem{Amendola:2012ys}
  L.~Amendola {\it et al.} (Euclid Theory Working Group Collaboration),
  Cosmology and fundamental physics with the Euclid satellite,
  Living Rev.\ Relativity  {\bf 16}, 6 (2013)\,.

\bibitem{Andre:2013afa}
  P.~Andre {\it et al.} (PRISM Collaboration),
  PRISM (polarized radiation imaging and spectroscopy mission): A white paper on the ultimate polarimetric spectro-imaging of the microwave and far-infrared sky,
  arXiv:1306.2259\,.

\bibitem{Brandenberger:2014faa}
  R.~H.~Brandenberger, A.~Nayeri, and S.~P.~Patil,
  Closed string thermodynamics and a blue tensor spectrum,''
  Phys.\ Rev.\ D {\bf 90}, 067301 (2014)\,.

\bibitem{Hwang:2005hb}
  J.~C.~Hwang and H.~Noh,
  Classical evolution and quantum generation in generalized gravity theories including string corrections and tachyon: Unified analyses,
  Phys.\ Rev.\ D {\bf 71}, 063536 (2005).

\bibitem{Kallosh:2013hoa}
  R.~Kallosh and A.~Linde,
  Universality class in conformal inflation,
  J. Cosmol. Astropart. Phys. {\bf 07} (2013) 002.

\end{thebibliography}\endgroup

\end{document}